\documentclass{scrartcl}
\usepackage[utf8]{inputenc}
\usepackage{units}
\usepackage{array}
\usepackage{graphicx}
\usepackage{epstopdf}
\usepackage{float}
\usepackage{amsmath}
\usepackage{amssymb}
\usepackage{esint}
\usepackage{multicol}
\usepackage{changepage}
\usepackage{textcomp}
\usepackage{hyperref}
\usepackage{xcolor}
\usepackage{float}
\usepackage{soul}
\usepackage{upgreek} 
\usepackage{authblk} 
\usepackage{cleveref}
\usepackage{placeins}
\usepackage{floatpag}

\newcommand{\blue}[1]{\textcolor{blue}{{#1}}}

\sethlcolor{yellow}

\title{Src kinase slows collective rotation of confined epithelial cell monolayers}
\author[1$\ast$]{Nastassia Pricoupenko}
\author[1$\ast$]{Flavia Marsigliesi}
\author[2]{Philippe Marcq}
\author[1]{Carles Blanch-Mercader}
\author[1]{Isabelle Bonnet}
\affil[1]{Institut Curie, Université PSL, Sorbonne Université, CNRS UMR168, Physics of Cells and Cancer, F-75005 Paris, France}
\affil[2]{Physique et Mécanique des Milieux Hétérogènes, PMMH, CNRS, ESPCI Paris, Université PSL, Sorbonne Université, Université Paris Cité, F-75005, Paris, France}
 \affil[$\ast$]{Equal contribution}

\date{}

\begin{document}

\maketitle
\section*{Abstract}
Collective cell migration is key during development, wound healing and metastasis and relies on coordinated cell behaviors at the group level. Src kinase is a key signalling protein for physiological functions of epithelia, as it regulates many cellular processes, including adhesion, motility, and mechanotransduction. Its over-activation is associated to cancer aggressiveness. Here, we take advantage of optogenetics to precisely control Src activation in time and show that its pathological-like activation slows collective rotation of epithelial cells confined into circular adhesive patches. We interpret velocity, force and stress data during period of non-activation and period of activation of Src thanks to an hydrodynamic description of the cell assembly as a polar active fluid. Src activation leads to a 2-fold decrease in the ratio of polar angle to friction, which could result from increased adhesiveness at the cell–substrate interface. Measuring internal stress allows us to show that active stresses are subdominant compared to traction forces. Our work reveals the importance of fine-tuning the level of Src activity for coordinated collective behaviors.

\noindent \textbf{Keywords:} epithelial cells, collective rotation, Src oncoprotein, optogenetics, TFM, BISM, active polar fluid

\section{Introduction}
Collective behaviors are important modes of cell migration \textit{in vivo}, be it throughout development or in diseases \cite{Roerth2009,Scarpa2016,Hakim2017}. Modes of collective cell migration are affected by different parameters which can be classified into two categories: the intrinsic cellular factors related to the biological state of the cells and the extracellular signals from the properties of the cells environment, including the confining geometry \cite{Ladoux2017}. The striking case of spontaneous collective \textit{rotation} has recently attracted substantial attention. For instance, the elongation of developing \textit{Drosophila} eggs is accompanied by coordinated rotation along the long axis of the ellipsoid \cite{Haigo2011}. During metamorphosis, male \textit{Drosophila} genitalia undergo collective rotation around the antero-posterior axis \cite{Kuranaga2011}. \textit{In vitro}, mammary epithelial cells self-organize into acini which may exhibit coherent angular motion \cite{Tanner2012}. Collective cell rotation occurs robustly in the case of micropatterned mammalian epithelial cells in a planar domain, confined either within a disk \cite{Doxzen2013}, within a square \cite{Peyret2019}, or within a ring \cite{Wan2011,Jain2020,LoVecchio2024}. Its establishment requires neither large cell numbers \cite{Huang2005} nor large confining domains \cite{Segerer2015}. Modeling efforts \cite{Hakim2017} attempting to ascertain which physical ingredients may be responsible for collective cell rotation include agent-based models \cite{Szabo2006,Li2014}, vertex models \cite{Sato2015}, active nematic \cite{Woodhouse2012,Ascione2023,Schimming2023} and active polar \cite{BlanchMercader2021a,BlanchMercader2021b} hydrodynamic descriptions.

\noindent Src, the protein product of the \textit{c-src} gene, is a non-receptor tyrosine kinase \cite{Yeatman2004,Ingley2008} that phosphorylates a variety of protein substrates that perform specific cellular functions \cite{Parson2004}. Src kinase exists in two states, active and inactive, enabling it to act as a signalling switch in response to extracellular signals. Src is expressed ubiquitously in all cell types and constitutes an important signaling protein, regulating a wide range of cellular processes such as motility, adhesion turnover and survival \cite{Westhoff2004}. As c-Src has been identified as an oncogene, its activity must be tightly regulated, since its misregulated activation leads to aberrant signalling which results in changes of cell phenotype. For instance, increased Src activity has been shown to reduce adhesion between cells \cite{Behrens1993}, to regulate actin dynamics \cite{Destaing2008} and to participate in epithelio-mesenchymal transition \cite{Avizienyte2005}. The combined effects of Src 
activation promote cell migration and may, in turn, facilitate invasion by increasing the metastatic potential of cells \cite{Irby1999}. Besides, several evidences implicate elevated expression and/or activity of Src in both cancer development and metastasis \cite{Talamonti1993,Frame2002}, suggesting that Src activity may play a role in tumor dissemination. It is therefore important to understand how deregulation of kinases in general, and Src in particular, affects the mechanical properties of cell assemblies. Several studies have already investigated the mechanical effect of Src activation on isolated cells \textit{in vitro}. Polackwich and colleagues showed that breast cancer cells overactivating Src spread more and exert higher traction forces than wild type cells in a 3D collagen environment \cite{Polackwich2013}. Karginov \textit{et al.} used a rapamycin-inducible Src to show that activation of Src led to cell spreading too as well as an increase in the number of focal adhesions in HeLa cells \cite{Karginov2014}. The behavior of assemblies of cells does not necessarily follow the same path and needs to be studied.

\noindent In this paper, we combine microfabrication, optogenetics, mechanical characterization, and theoretical modeling to investigate how the activation of Src kinase affects the collective rotation of epithelial cell colonies. We choose \textit{in vitro} approaches that offer a close control of experimental conditions. In particular, 
microfabrication allows us to confine 2D epithelial cell colonies in disks.
To precisely control the timing of Src activation, we use a stable optogenetic epithelial cell line in which the oncogenic Src signalling is activated upon blue-light exposure \cite{Moitrier2019,Kerjouan2021}. The mechanical state of the monolayers is characterized by measuring cell flows, cell traction forces as well as the internal stresses either in the normal state (in the dark, when Src is not overactivated) or when Src is overactivated (under blue-light illumination). We observe that photoactivation of Src alters collective rotation by decreasing azimuthal velocity. We use a previously-developed active polar fluid model to describe the spontaneous rotation of a circularly confined cell sheet \cite{BlanchMercader2021b,Blanch2017effective}. We first show that traction forces are the main ingredient contributing to rotation while active stresses (whether chiral or not) do not significantly contribute to rotation. Comparing theoretical predictions with experiments reveals that Src activation is associated with a decrease in the ratio of polar angle to friction. 

\section{Results}
\subsection{Experimental strategy.} 
\begin{figure}
\centering
	\includegraphics[width=\linewidth]{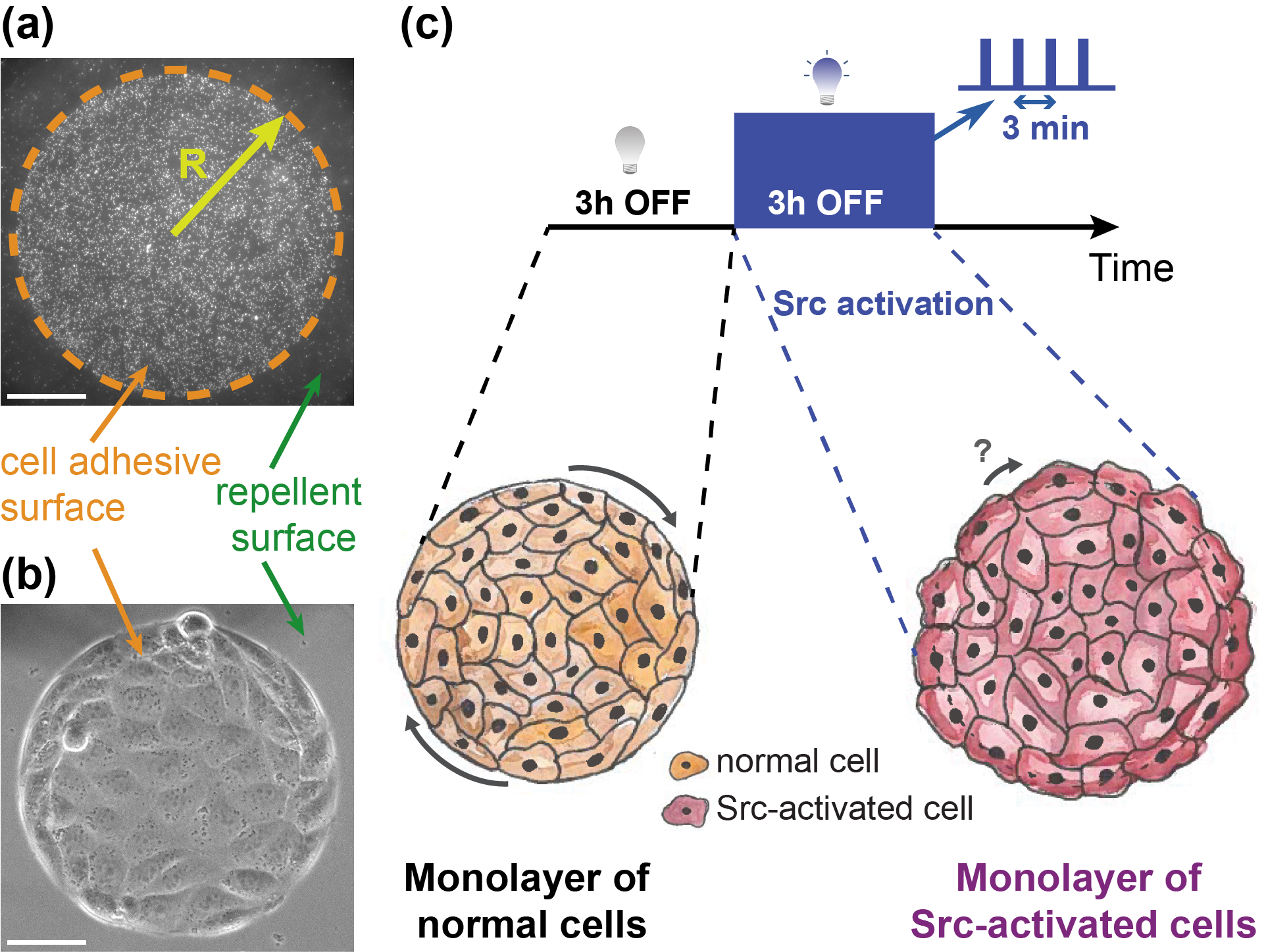}
\caption{Experimental setup to study the impact of Src activation on collective rotation. (a) Fluorescent image of the micron-sized beads used for computing traction forces, mainly located inside the circular pattern of radius $R$. (b) Phase-contrast image of MDCK optoSrc cells confined in 100~$\upmu$m radius circular pattern, same location as in a. Scale bars: 50~$\upmu$m. (c) Blue-light illumination sequence used to transiently activate Src kinase (OFF-ON case): 3\,hrs OFF then 3\,hrs ON. Under blue-light illumination (which consists in 200-ms blue-light pulses separated by 3\,min), MDCK optoSrc cells activate Src (in pink in the schematic) while they behave like MDCK normal cells in the dark (in orange in the schematic).}
\label{fig_setup}
\end{figure}

\noindent We benefited from a light-sensitive stable epithelial cell line, the MDCK (Madin Darby Canine Kidney) optoSrc cells \cite{Moitrier2019,Kerjouan2021}. When exposed to blue light, these cells overactivate the protein Src, which is known to be overexpressed in many cancers \cite{Irby2000,Wheeler2009}. Otherwise, MDCK optoSrc cells keep a normal phenotype. In addition to being highly specific and non-invasive, our optogenetic system offers a precise control in time of the activation of the Src oncoprotein. \\

\noindent We have previously shown \cite{Moitrier2019b} that free islands of light-activated MDCK optoSrc cells behave differently from MDCK normal cells. When exposed to blue light, clusters of MDCK optoSrc cells quickly spread out and eventually lose cell-cell contacts. After a few hours, groups of MDCK optoSrc cells have completely scattered while groups of MDCK normal cells remain cohesive even under similar blue-light stimulation. To avoid this, we decided to confine the cells, even though confinement is not necessary to observe the rotational motion of adherent cells \cite{Ascione2023}. We have chosen a circular confinement of radius $R = 100$\,$\upmu$m. Under such a confinement, collective rotation is expected for MDCK normal cells \cite{Doxzen2013,Deforet2014}, which will subsequently constitute our reference for collective motion. \\

\noindent Confined monolayers of MDCK optoSrc cells were grown on polyacrylamide gels of Young modulus $\sim$40\,kPa loaded with fluorescent beads to measure traction forces (Fig.~\ref{fig_setup}a). These elastic gels are patterned with fibronectin-coated circular patches of radius $R = 100\,\upmu$m to confine the monolayer (Fig.~\ref{fig_setup}b, Fig.~S1). \\

\noindent Experiments start when the cells are at \textit{confluence}, \textit{i.e.} when there are no holes in the cell monolayer. As expected, MDCK optoSrc cells confined in 100\,$\upmu$m radius disks and not exposed to blue light have the same phenotype as MDCK normal cells: they exhibit collective rotation (Movie~1). We measured the duration of these collective rotations at approximately (7$\pm$4)\,hrs (sd, N=19, Fig.~S2a), in agreement with what has already been reported for MDCK normal cells \cite{Deforet2014}. To study the impact of Src activation on this collective motion, we therefore chose to focus on a 6-hr period: for the first 3 hrs, MDCK optoSrc cells are kept in the dark, then subjected to blue light for the following 3 hrs: we call this time sequence the OFF-ON case, as shown in Fig.~\ref{fig_setup}c. Our control experiment, called the OFF-OFF case, consists in keeping MDCK optoSrc cells for 6~hrs in the dark. 

\subsection{Src activation slows down collective rotation.} 
\begin{figure}[!ht]
\centering
	\includegraphics[width=\linewidth]{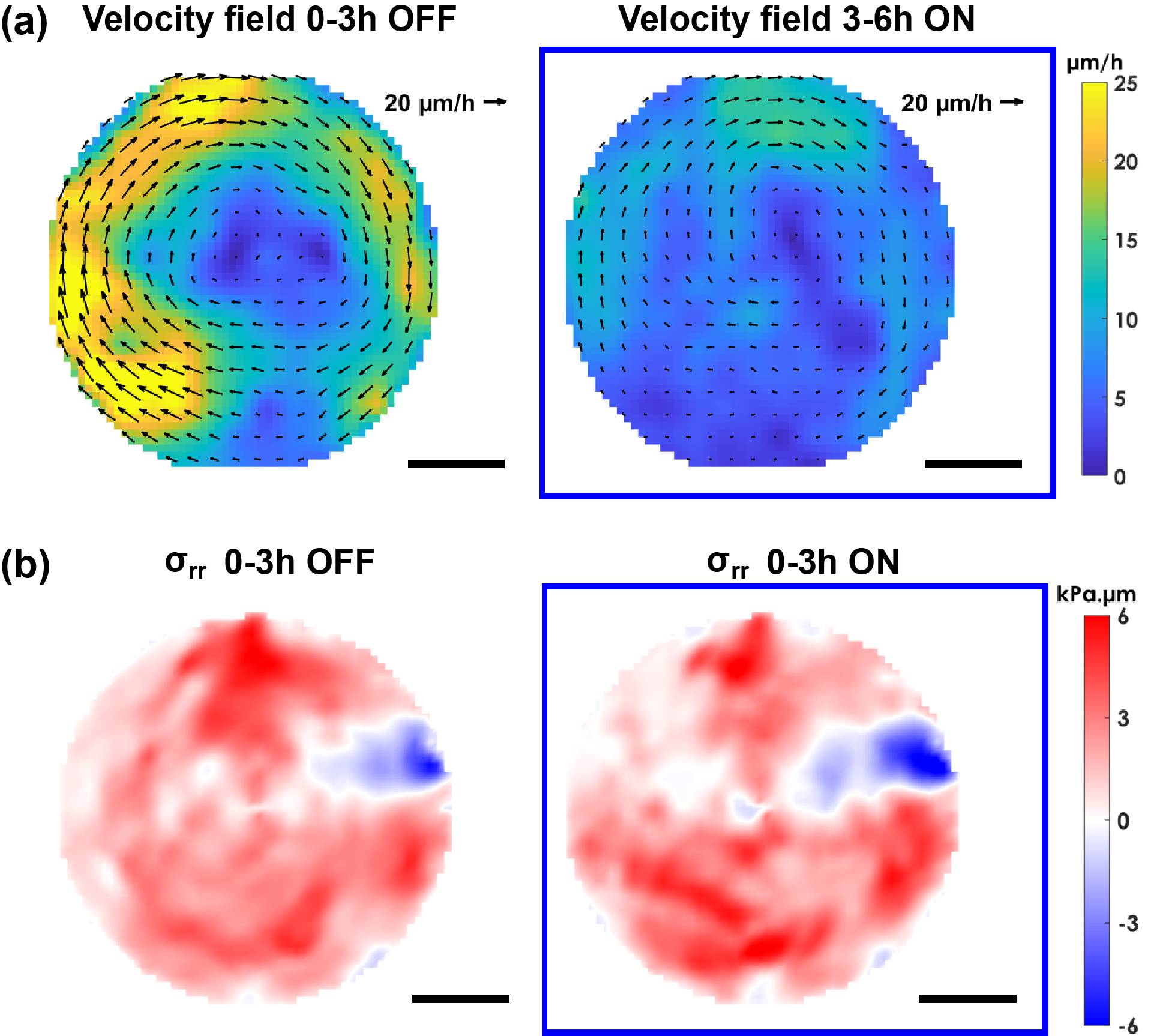}
\caption{When Src is activated, collective rotation is slowed down. (a) Time-averaged velocity magnitude contour plot superimposed with velocity vector plot obtained by PIV for the first 3\,hrs (left, OFF) and for the last 3\,hrs (right, ON, Src is activated) for a given colony. Only one arrow out of three is drawn for visibility. (b) Time-averaged stress component $\sigma_{rr}$ measured by BISM over the first 3\,hrs (left, OFF) and over the last 3\,hrs (right, ON, Src is activated), same colony as in a. The colony is under tension, except for a small domain appearing in blue. Note, that, in 2d, stresses are expressed in kPa$\cdot\upmu$m. Scale bars: 50~$\upmu$m. See Movie~2 for the spatio-temporal dynamics of the velocity field and the stress field.}
\label{fig_result}
\end{figure}

\noindent Qualitatively, during the OFF phase, the optoSrc cells behave like normal cells. When Src activation starts (ON phase), cells spread out with the formation of lamellipodia as previously observed \cite{Karginov2014}, causing Src-activated cells, located at the periphery, to move outside of the adhesive pattern (Movie~2). We thus estimated the amplitude of monolayer radius fluctuation $\frac{\Delta R}{R} \sim 10\%$ (Fig.~S2b).\\

\noindent For quantitative analyses, we used Particle Image Velocimetry (PIV, \cite{Petitjean2010,Vig2016}) to measure cell displacements within the monolayer and Traction Force Microscopy (TFM, \cite{Style2014}) to measure the traction forces that cells exert on the substrate. Finally, Bayesian Inference Stress Microscopy (BISM, \cite{Nier2016}) was used to infer intracellular stresses within the cell sheet from traction force data (see Methods). Circular geometry encourages us to decompose traction force and velocity fields into radial and azimuthal components, $\vec{e}_r$ and $\vec{e}_\theta$ denoting the corresponding unit vectors. Furthermore, rotational invariance, observed in a statistical sense, makes it possible to only consider the radial dependence of the quantities. Since the amplitude of the radius fluctuations is small ($\frac{\Delta R}{R} \sim 10\%$) and the density of beads outside the pattern is too low for the traction forces to be properly computed as shown in Fig.~\ref{fig_setup}a, our analyses were carried out on the data located inside the initial adhesive pattern: $r\leq R$. \\

\noindent We first compared the mechanical state of the monolayer between the OFF and ON cases by time-averaging velocity, force and stress components over the first 3\,hrs (OFF) and over the last 3\,hrs (ON). During the OFF period, confined optoSrc cells undergo a collective rotation, similar to that observed for normal cells: the radial velocity component is relatively small throughout the pattern while the azimuthal one increases with\,$r$. When Src is activated, $v_r$ increases a little translating the spreading of cells out of the adhesive pattern while the amplitude of the azimuthal component is significantly reduced (Fig.~\ref{fig_result}a). The time evolution of velocity magnitude, averaged over the whole pattern, confirms that the monolayer speed is significantly decreased when Src is activated, which is not the case when kept in the dark (Fig.~S3a-c).  Cell density increase over time could be responsible for the observed slowdown \cite{Abercrombie1953}. We therefore compared the time evolution of cell density between the two types of experiment (control and Src activation). Over the course of 6\,hrs, cell density increases (Fig.~S4a-b). However, there is no significant difference for initial ($t=0$) and final ($t=6$\,hr) cell densities in the two cases (Fig.~S4c), suggesting that the slowdown observed the last 3\,hrs is not due to the increase in cell density but associated with Src activation.  \\

\noindent Although slightly larger in norm when Src is activated, traction force maps do not show significant differences between OFF and ON cases after temporal averaging (Fig.~S5a). Components of the stress tensors, such as the normal stress in the radial direction, $\sigma_{rr}$, do not show significant differences between OFF and ON cases (Fig.~\ref{fig_result}b) either. The isotropic stress $\sigma_{\mathrm{iso}} = (\sigma_{rr} + \sigma_{\theta \theta})/2$ is positive on average, revealing a tensile system (Fig.~S5b) expected of confined epithelial cell monolayers \cite{Peyret2019,Nier2016}. The amplitude of the deviatoric component $(\sigma_{rr} - \sigma_{\theta \theta})/2$ is almost zero except close to the margin (Fig.~A1c). The shear component $\sigma_{r \theta}$ is at least one order of magnitude smaller than that of the isotropic stress, as has also been observed in epithelial cell monolayers confined in square domains \cite{Nier2016b}, see Fig.~4f. Statistical tests performed on the average physical variables confirm our claims (Fig.~S6).

\section{Theoretical description}
\label{sec:3}
\noindent Here, we briefly summarise the physical model 
of an active polar fluid similar to one previously used~\cite{BlanchMercader2021b,Blanch2017effective} to describe the collective rotation of cell monolayers when confined to a disc. For a more a detailed derivation of the equations below and further justifications of the theoretical arguments, we refer the reader to~\cite{BlanchMercader2021b} and the Electronic Supplementary Information. \\

\noindent The physical description is in 2d and the thickness $h$ of the layer is considered uniform, for simplicity. The 2d stress $\sigma$ thus reads $\sigma = \sigma_{\mathrm{3d}} \times h$, where the 3d stress $\sigma_{\mathrm{3d}}$ is expressed in Pascal. $\sigma$ is thus expressed in Pa$\cdot$m. Cell monolayers are characterised by the polarity field $\mathbf{p}$ that accounts for the coarse-grained cell polarity and by the velocity field $\mathbf{v}$ that accounts for the coarse-grained cell velocity. The geometry of the system is a disc of radius $R$. The study focuses on steady-state configurations that are independent on the azimuthal coordinate $\theta$, and only depend on the radial coordinate $r$. In polar coordinates, the polarity field can be expressed as  $\mathbf{p}=S(r)(\cos(\psi),\sin(\psi))$, where $S(r)$ corresponds to the polar order parameter and $\psi$ is the angle between the polarity field and the radial direction, which for simplicity, is considered uniform. Assuming that the dynamics of the polarity field relax fast and ignoring alignment by flows, the dynamic equations for the polarity field becomes
\begin{eqnarray}
0&=&-\chi S(r)+\kappa\left(\partial_{rr}S(r)+\frac{\partial_r S(r)}{r}-\frac{S(r)}{r^2}\right)
\end{eqnarray}
where $\chi$ is the elastic constant associated with the variations of polarity amplitude $S$ and $\kappa$ is the Frank constant. Their ratio gives rise to a lengthscale $L_c=\sqrt{\kappa/\chi}$ that controls the penetration length of boundary-induced polar order. \\

\noindent In experiments, we computed the spatially-averaged velocity divergence $<\mathbf{\nabla}\cdot\mathbf{v}>_{r,\theta}$ as a function of time for each colony (Fig.~S3d). The velocity divergence peaks right after the blue-light is turned ON, in line with the radial displacement of cells that is observed at this moment (Movie~2). Since the average velocity divergence during the ON phase ($0.01$\,hr$^{-1}$) is smaller than the typical frequency rotation ($\mathrm{max}(v_{\theta})/R=0.1$\,hr$^{-1}$), we consider here the incompressible limit $\mathbf{\nabla}\cdot\mathbf{v}=0$, which is different with our previous works \cite{Moitrier2019,Blanch2017effective}, but similar to, \textit{e.g.} \cite{banerjee2015propagating,Ascione2023}. In addition, the effects associated to cell density variations in time and space are ignored. Assuming that the velocity components only depend on the radial coordinate, and that there is no flux at the boundaries, the radial component of the velocity field must vanish everywhere as a consequence: $v_r=0$. \\

\noindent In the absence of inertial effects, the momentum conservation equation reduces to the force balance equation, which reads
\begin{eqnarray}
\partial_r\sigma_{rr}+\frac{\sigma_{rr}-\sigma_{\theta\theta}}{r}&=&T_{r}\\
\partial_r\sigma_{\theta r}+\frac{\sigma_{r\theta}+\sigma_{\theta r}}{r}&=&T_{\theta}
\end{eqnarray}

\noindent The gradients of total 2d stresses $\sigma$ are balanced by the traction forces $\mathbf{T}$ that arise from interaction forces between cells and the underlying substrate. The component of the total stress tensor are
\begin{align}
\sigma_{rr,\theta\theta}&=-P\label{eq:StressTensorRR}  \\
\sigma_{r\theta,\theta r}&=\eta\left(\partial_r v_{\theta}(r)-\frac{v_\theta(r)}{r}\right)\label{eq:devStressTensorRTheta} 
\end{align}
The total 2d stresses $\sigma$ include only the diagonal part of the Ericksen stresses (see ESI). Recent work \cite{Tlili2020migrating} showed that spreading MDCK cell monolayers behave like a Maxwell material with a viscoelastic timescale of $~1$~hr. On the time scale of this experiment, we consider that the cell monolayer behaves as a viscous fluid, and include in the total 2d stresses $\sigma$ viscous stresses proportional to the shear viscosity $\eta$. 
Because the system is incompressible, the pressure field $P$ acts as a Lagrange multiplier to ensure the constraint that the cell number density is uniform.
The traction forces that arise from the interactions between cells and the substrate 
\begin{align}
T_{r}&=-T_0 S(r)\cos(\psi)\label{eq:TractionR}  \\
T_{\theta}&=\xi v_\theta-T_0 S(r)\sin(\psi)  \label{eq:TractionO} 
\end{align}
include two contributions that is a viscous friction proportional to the velocity field $\mathbf{v}$, with a friction coefficient $\xi$, and active traction forces proportional to the polarity field $\mathbf{p}$, with an amplitude $T_0$. Note that the radial component of viscous friction vanishes because $v_r=0$. In this convention, $\mathbf{T}$ represents the force surface densities that cells exert on to the substrate, measured in Pa.\\

\begin{figure}[!h]
\centering
	\includegraphics[width=0.75\linewidth]{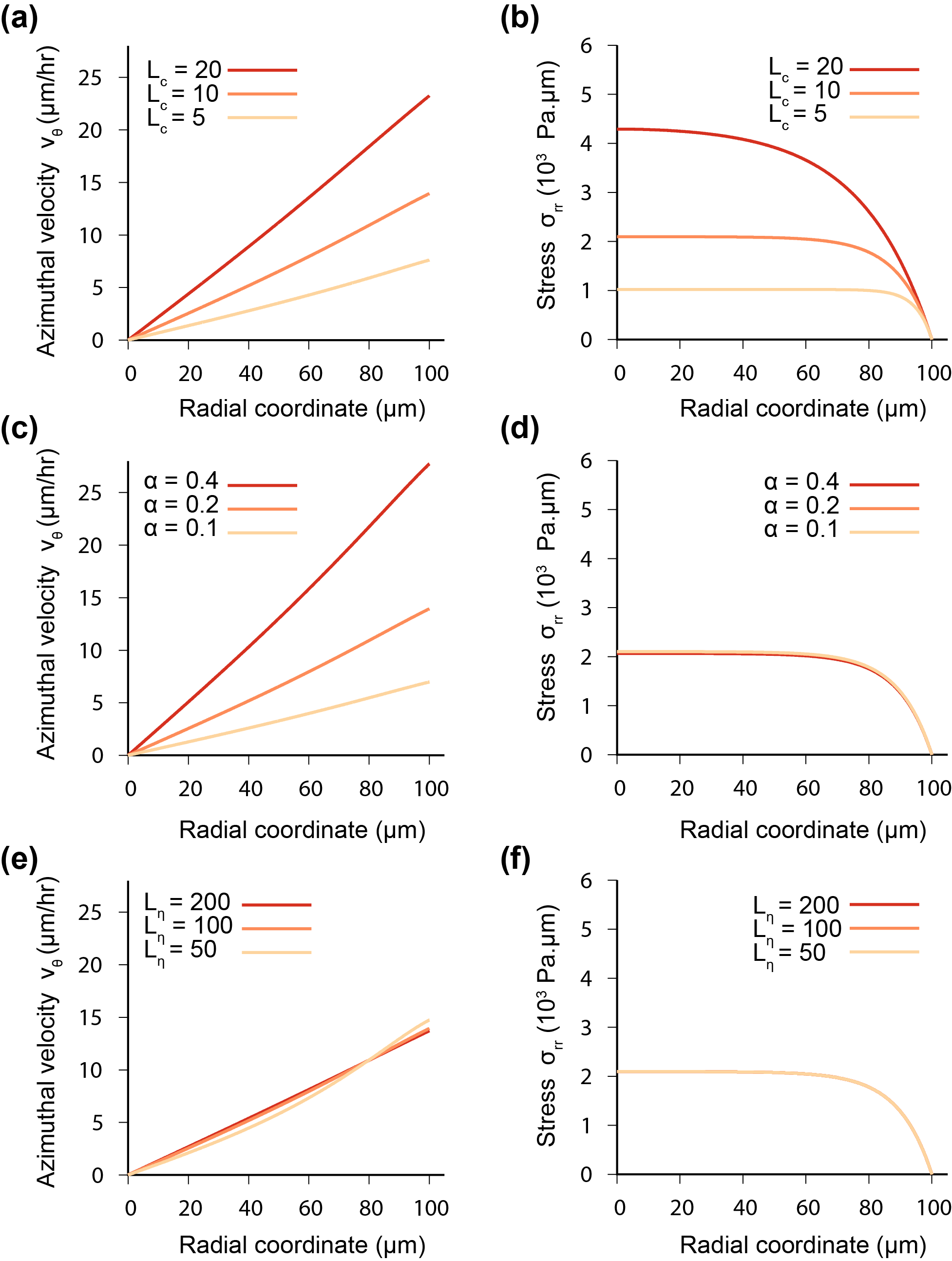}
\caption{Dependence of velocity and stress radial profiles on material parameters. The left panels correspond to the azimuthal velocity profile $v_\theta(r)$~(Eq.~\eqref{eq:velO}), and the right panels to the stress $\sigma_{rr}(r)$ profile (Eq.~\eqref{eq:SigmaRR}). From top to bottom, the parameters $L_c$, $\alpha$ or $L_\eta$ have been doubled (dark orange) or halved (light orange). To vary $\alpha=\sin(\psi)/\xi$, the orientation $\psi$ is fixed and $\xi$ 
 varies. The red curves is for the reference parameter set: $T_0=200$~Pa, $L_c=10~\upmu$m, $L_\eta=100~\upmu$m, $\xi=0.5~$Pa$\cdot$hr/$\upmu$m, $\psi=0.1$~rad ($\alpha=0.2$~$\upmu$m/Pa$\cdot$hr) and $R = 100~\upmu$m. In legends, the lengthscales $L_\eta$ and $L_c$ are in $\upmu$m and $\alpha$ is in~$\upmu$m/Pa$\cdot$hr.}
\label{fig_phase_diagram}
\end{figure}

\noindent At the boundaries of the circular confinement $r=R$, we impose free-boundary conditions that is $\sigma_{rr}|_{r=R}=\sigma_{\theta r}|_{r=R}=0$. In addition, we impose boundary-induced polar order and without loss of generality, the polar order scale at the interface is set to its maximal value $S(R)=1$.\\

\noindent The previous ODEs have an explicit analytical solution that takes the form: 
\begin{eqnarray}
S(r)&=&\frac{I_1(\frac{r}{L_c})}{I_1(\frac{R}{L_c})}\\
v_\theta(r)&=&\frac{T_oL_c\sin(\psi) }{\xi(L_c^2-L_\eta^2)}\frac{L_c I_1(\frac{r}{L_c})I_2(\frac{R}{L_\eta})-L_\eta I_1(\frac{r}{L_\eta})I_2(\frac{R}{L_c})}{I_1(\frac{R}{L_c})I_2(\frac{R}{L_\eta})}\label{eq:velO}\\
\sigma_{rr}(r)&=&-P(r)=T_oL_c\cos(\psi) \frac{I_0(\frac{R}{L_c})-I_0(\frac{r}{L_c})}{I_1(\frac{R}{L_c})}\label{eq:SigmaRR}
\end{eqnarray}

\noindent where $I_n(x)$ is a modified Bessel function of the first kind of order 
$n$ and $L_{\eta}=\sqrt{\eta/\xi}$ is the friction length. Similarly, one can find expressions for the other components of the stresses and the traction forces (see ESI). These solutions were used in the fitting procedure detailed below. \\

\noindent We find that the above expressions depend on the two lengthscales $L_c$ and $L_\eta$. In addition, the scale of the stress component is set by $T_oL_c\cos(\psi)$, and the scale of the velocity field is set by $T_o\alpha$, where $\alpha=\sin(\psi)/\xi$ is a motility coefficient, in the sense that it quantifies the propensity of the system to rotate. If $\alpha = 0$, which corresponds to polarity purely radial or infinite friction, then there is no rotation. Fig.~\ref{fig_phase_diagram} shows the dependencies of the azimuthal component of the velocity and the radial stress profiles on these three parameters $L_c$, $L_\eta$, and $\alpha$. In this case, the mechanical profiles mainly depend on the parameters $\alpha$ and $L_c$. By varying $L_c$, the slope of the velocity profile and the scale of the stresses can be modulated, whereas varying the coefficient $\alpha$ mainly influences the slope of the velocity profile. In the parameter regime $L_\eta\geq R$, the energy dissipation is dominated by viscous stress and variations in $L_\eta$ weakly modify the velocity and the stress profiles. Because the velocity profiles are approximately linear in the radial coordinate, the shear flows vanish, and viscous friction is necessary to ensure that the total torque from the traction forces $\mathbf{T}$ vanishes too.   \\

\section{Characterization of MDCK optoSrc cell monolayers}\label{sec:3.1}
We use the framework developed above to analyse the collective rotation of MDCK optoSrc cells. To determine the tissue parameters, we analyse the radial profiles of the coarse-grained physical fields, including the cell velocity, the traction forces and the mechanical stresses. The fitting procedure is detailed in Section~\ref{sec:3.2} and the results are presented in Section~\ref{sec:3.3}. 

\subsection{Fitting procedure.}
\label{sec:3.2}
\noindent Here, we summarise the fitting procedure used in the analysis. The difference between the numerical and experimental fields are quantified \textit{via} an error function ${\cal E}$ that is given below. The parameter set that gives the minimal error ${\cal E}_{\mathrm{min}}$ then provides the sought material parameters. Confidence intervals are  determined by considering the range of parameter values that yields an error within $20\%$ of the minimal error, that is, for which ${\cal E} < 1.2~{\cal E}_{\mathrm{min}}$.\\

\noindent Experimental radial profiles, denoted with the superscript $e$, were averaged over time and over colonies.
The theoretical profiles were obtained as explained in the previous section. These profiles depend on the following parameters: two material lengthscales $L_c$ and $L_\eta$, the traction scale $T_0$, the friction coefficient $\xi$, the polarity orientation $\psi$ and the radius of the confinement $R$. In experiments, the radius of confinement is fixed, and therefore we set $R=100$~$\upmu$m. This leaves us with five parameters to fit six radial profiles: two components of the velocity field ($v_r,v_\theta$), two components of the traction force density ($T_r,T_\theta$), and two components of the stresses ($\sigma_{rr},\sigma_{r\theta}$). Solutions were computed within the parameter space $(L_c,L_\eta,\psi,\xi,T_0,)=(1,10^2)\times(10^{-3},10^3)\times(\pi/2*10^{-2},\pi/2)\times(10^{-3},10^3)\times(10,10^4)$, where the unit set is given in Pascal, micrometers, hours and radians. \\

\noindent The error function is defined as:
\begin{align}
{\cal E}&=&\sum_{\beta=r,\theta}\left(\sum_i\frac{|v_{\beta,i}^e-v_{\beta,i}|}{v_0}+\sum_i\frac{|T_{\beta,i}^e-T_{\beta,i}|}{t_0}+\sum_i\frac{|\sigma_{r\beta,i}^e-\sigma_{r\beta,i}|}{\sigma_0}\right)\label{eq:error}
\end{align}
where $|\,|$ denotes the L1-norm. Here, the index $i$ indicates that samples are taken at discrete radial positions $r_i$ with a spacing of $2.5~\upmu$m. The numbers $v_0=\max_i(|v^e_{r,i}|,|v^e_{\theta,i}|)$, $t_0=\max_i(|T^e_{r,i}|,|T^e_{\theta,i}|)$, and $\sigma_0=\max_i(|\sigma^e_{rr,i}|,|\sigma^e_{r\theta,i}|)$ are used to normalise and adimensionalise components of the error function.

\subsection{Results.}\label{sec:3.3}
First, we applied the above fitting procedure to cell monolayers in the OFF-ON case. 
Theoretical fits are shown in Fig.~\ref{fig_4} and the value of the fitted parameters are summarised in Table~\ref{tab:table1}. In appendix~B, we applied a similar procedure to cell monolayers in the OFF-OFF case and summarised the fitted parameters in Table~\ref{tab:table1} too. \\

\begin{figure}[!h]
\centering
	\includegraphics[width=0.75\linewidth]{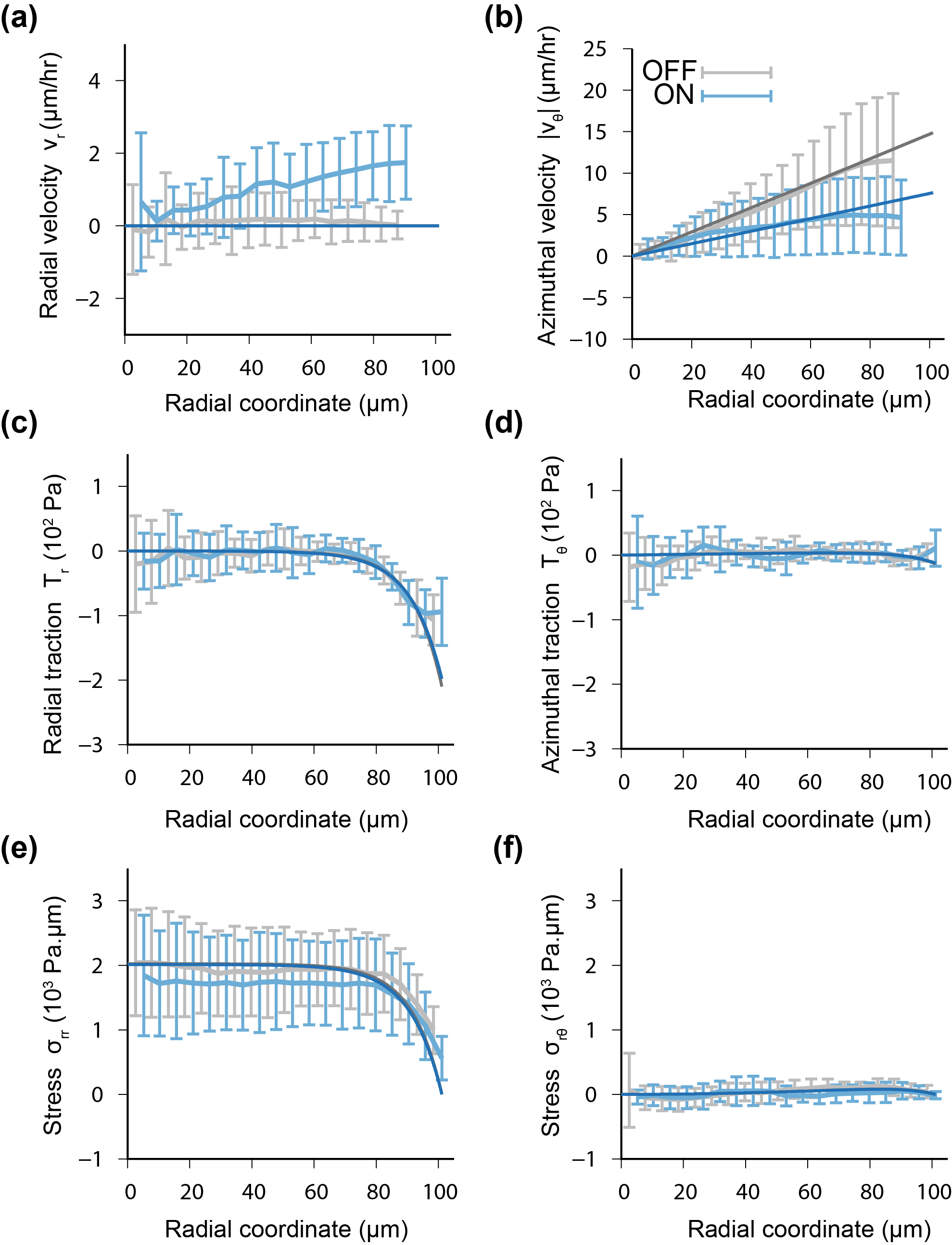}
\caption{Comparison of experimental profiles in the OFF-ON case with theoretical fits. OFF data are in gray while ON data are in blue. (a) and (b) Radial profiles of the radial and azimuthal components of cell averaged velocity field $v_{r}$ and $v_{\theta}$, respectively. Note the different scales used on the $y$-axes. (c) and (d) Radial profiles of the radial and azimuthal components of cell averaged traction force field $T_{r}$ and $T_{\theta}$, respectively. (e) and (f) Radial profile of the components of cell averaged stress $\sigma_{rr}$ and $\sigma_{r\theta}$, respectively. Plots correspond to data averaged over $N=17$ colonies and error bars correspond to standard deviation after azimuthal and temporal averaging. Theoretical fits are shown as solid curves in dark gray for the 3\,hrs OFF and in blue for the 3\,hrs ON. Fitted parameter values are in Table~\ref{tab:table1}.}
\label{fig_4}
\end{figure}

\noindent Upon blue-light illumination, collective cell rotation slowed, while the average stresses and traction forces remained unchanged. The averaged velocity in the azimuthal direction decreases from 3\,hrs OFF to 3\,hrs ON period, while in the radial direction, we observed an increase in amplitude (Fig.~\ref{fig_4}a-b). The radial traction force was negative and minimal at the confinement interface, whereas the azimuthal component vanished (Fig.~\ref{fig_4}c-d). Concurrently, we found that the stress $\sigma_{rr}$ was positive and plateaued near the center, indicating a tensile component, while the component $\sigma_{r\theta}$ vanished (Fig.~\ref{fig_4}e-f). \\

\noindent When fitting the experimental curves for the OFF-ON experiments, most of the parameters remain unchanged except for the value of the motility coefficient $\alpha$. This parameter mainly controls the scale of the rotational flow velocity (Fig.~\ref{fig_phase_diagram}c-d), and it is significantly decreased from $0.22\pm 0.03~\upmu$m/Pa$\cdot$hr in the 0-3\,hr OFF period to $0.12\pm 0.01~\upmu$m/Pa$\cdot$hr in the 3-6\,hr ON period. When fitting the experimental curves for the OFF-OFF experiments, we found that all parameters remain unchanged, including  $\alpha=0.14\pm0.02~\upmu$m/Pa$\cdot$hr (Table~\ref{tab:table1}). In addition, across experiments, the polarity length $L_c$ was of $10$~$\upmu$m, and the traction force amplitude $T_0$ varied in the range of $200-300$~Pa. Some material parameters were undetermined, specifically the friction length $L_\eta$, which had a lower bound on the range of $30$~$\upmu$m, and the polarity phase $|\psi|$, which had an upper bound of $0.3$~rad. In combination, these results showed that Src activation mainly influences the motility coefficient $\alpha$, which depends on both the orientation of the polarity field $\psi$ and the friction coefficient $\xi$. Besides, we checked that in the absence of a polarity field ($S(r)=0$), our theoretical description is unable to reproduce the experimental radial profiles.\\

\begin{table*}[h]
\small
\caption{Parameter sets obtained for MDCK optoSrc cell monolayers in the OFF-ON case with $N=17$ (first and second rows), and for cell monolayers in the OFF-OFF case with $N=19$ (third and fourth rows) - error bars correspond to standard deviation of all parameter values with ${\cal E}<1.2~{\cal E}_{\mathrm{min}}$. The bound of a fitting parameter means that a theoretical description with values of fitting parameter beyond this bound could not fit the experimental radial profiles.} \label{tab:table1}	
\begin{tabular*}{\textwidth}{@{\extracolsep{\fill}}llllll}
\hline
 Parameter & $L_c$ ($\upmu$m) & $L_\eta$ ($\upmu$m) & $T_0$ (Pa) & $|\psi|$ (rad)  & $\alpha$ ($\upmu$m/Pa$\cdot$hr) \\
\hline
 0-3h OFF case & $9\pm 3$ &  $>30$ &  $210\pm 60$ &  $<0.3$ &  $0.22\pm 0.03$   \\ 
\blue{3-6h ON case} & \blue{$10\pm 3$} &  \blue{$>40$} &  \blue{$200\pm 60$} &  \blue{$<0.4$} &  \blue{$0.12\pm 0.01$}   \\ \hline
 0-3h OFF case & $8\pm 3$ &  $>20$ & $270\pm90$  &  $<0.5$ &  $0.10\pm0.01$   \\ 
  3-6h OFF case & $7\pm 3$ &  $>30$ &  $320\pm 160$ &  $<0.3$ &  $0.14\pm 0.02$   \\ 
\end{tabular*}
\end{table*}

\section{Discussion}

\begin{figure}
 \centering
 	\includegraphics[width=\linewidth]{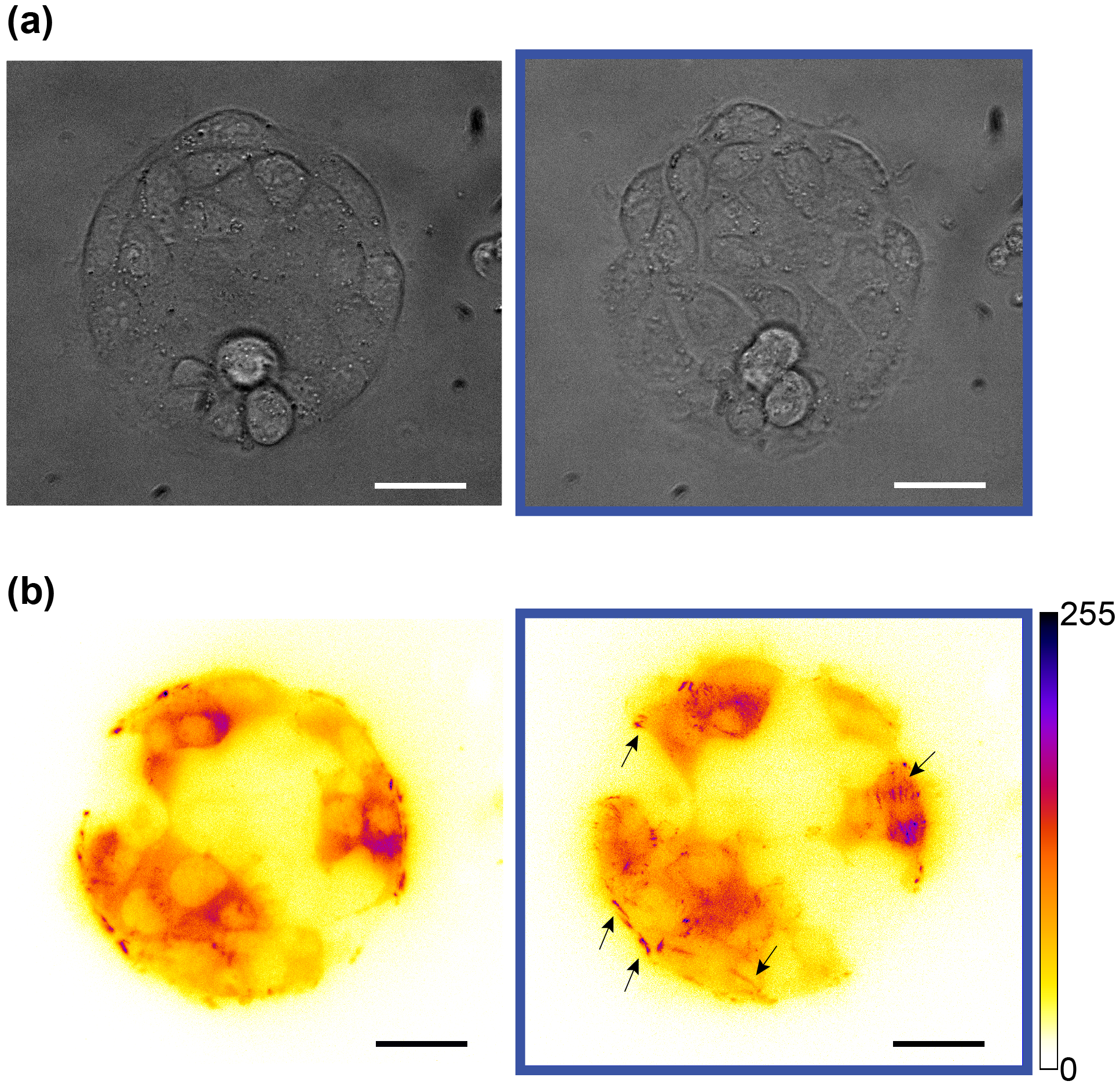}
\caption{When Src is activated, focal adhesions are more numerous. Representative view of MDKC optoSrc vinculin-iRFP cells confined in a circular pattern before (left) and after 1\,hr of blue-light illumination (right). (a) Phase-contrast images. (b) Epifluorescence images of vinculin-iRFP. Cells do not homogeneously express vinculin-iRFP, resulting in a heterogeneous fluorescent signal. Some focal adhesions are indicated by arrows. Scale bars: 25~$\upmu$m. }
 \label{fig_FA_labelling}
 \end{figure}

\noindent We studied here the spontaneous rotation of confined epithelial tissue. Mechanisms for collective rotation of active fluids driven by gradients of active stresses have been proposed \cite{Kruse2004asters,Furthauer2012taylor,Furthauer2013active,Hoffmann2020chiral}, specifically, achiral and chiral active stresses, which can represent different modes of coarse-grained active cell-cell and cell-substrate interactions. To explore the influence of these active mechanisms on collective rotation of cell monolayers, we analysed the stress and traction force components according to the direction of rotation in Appendix~A. By comparing these profiles between cell monolayers that rotated in the clock-wise and cell monolayers that rotated in the counter-clockwise direction, we conclude that the chiral active stresses are negligible. Besides, we did not find a significant difference between clockwise and counter clockwise direction in our experiments. The experimental profiles of the stress component  $\sigma_{rr}-\sigma_{\theta\theta}$ reveals that a contribution of the achiral active stress of at most $\zeta\Delta\mu \sim 10^3$~Pa$\cdot\upmu$m may be present in the system (see Appendix~A). Because the dimensionless ratio between the amplitudes of the two active processes is at most $\zeta\Delta\mu/R T_0\sim0.05$, we concluded that achiral active stresses are subdominant in front of the traction forces.\\

\noindent Combining our observations with the results presented in Table~1 yields additional quantitative information on the mechanical properties of MDCK cell monolayers. As seen in Fig.~\ref{fig_4}f, the shear stress is at most of the order of $10^2$ \, $\mathrm{Pa}.\upmu\mathrm{m}$. 
Eq.~\eqref{eq:devStressTensorRTheta} translates into the scaling relationship $\hat{\sigma}_{r\theta} \simeq \eta \hat{v}_\theta/R$, where $\hat{\sigma}_{r\theta}$ and $\hat{v}_\theta$ denote the orders of magnitude of the shear stress and azimuthal velocity, respectively. Since $\hat{\sigma}_{r\theta} \le 10^2$ \,$ \mathrm{Pa}.\upmu\mathrm{m}$ and $\hat{v}_\theta = 10$ \, $\upmu\mathrm{m\,hr}^{-1}$ (Fig.~\ref{fig_4}a), we deduce an upper bound for the 2d shear viscosity $\eta < 10^3\,\mathrm{Pa.h.\upmu m}$. This bound is consistent with an estimate of $\eta$ obtained for suspended cell monolayers in \cite{Harris2012} and for spreading cell monolayers \cite{Blanch2017effective}, using a typical monolayer height $h = 10\,\upmu$m. Since $L_\eta > 30\,\upmu$m, we next deduce an upper bound for the 2d friction coefficient $\xi < 1\,\mathrm{Pa .h.\upmu m^{-1}}$. Using $|\Psi|<0.2$ and $\alpha = 0.14$ (Table 1, OFF-OFF case), we obtain independently a similar bound $\xi < 1.4$\,$\mathrm{Pa .h.\upmu m^{-1}}$. Of note, these bounds are consistent with an estimate of $\xi$ obtained in \cite{Cochet-Escartin2014}.\\ 

\noindent Activation of Src kinase slows down collective rotation of confined epithelial tissue. Our model shows that a reduction in the motility coefficient $\alpha = \sin(\psi)/\xi$ can explain this disruption. The reduction in $\alpha$ may result from a decrease in $\sin(\psi)$ or an increase in $\xi$, or a combination of both. Analysis of the effect of Src on the angle $\psi$ was not conclusive (data not shown). Besides, we did not find difference for average value of traction forces $T_0$ between the OFF and ON cases, which is consistent with a constant amplitude of polarity. Consequently, the decrease of rotation velocity could not be attributed to a reduction in traction forces. Friction, which results from the turnover dynamics of cell adhesive bonds with the substrate, increases with the density of adhesive components. To examine the cell-substrate adhesive bonds during Src activation, we used an MDCK cell line that stably co-expresses optoSrc and vinculin fused to a near-infrared fluorescent protein (vinculin-iRFP). Vinculin is a major mechano-sensitive focal adhesion component \cite{Atherton2016}, and near-infrared imaging prevents optoSrc photoactivation. Vinculin-iRFP imaging reveals that MDCK optoSrc cells form focal adhesions that exhibit a classic arrowhead shape, but are more numerous upon Src activation, particularly in newly formed lamellipodia (Fig.~\ref{fig_FA_labelling}). Quantification confirms that focal adhesions are more numerous upon Src activation (Fig.~S7). 
Assuming that the vinculin fluorescence signal is a proxy for mechanically-engaged focal adhesion structures, we conclude that Src activation is associated with an increase of adhesive bonds with the substrate, and hence of friction. This observation leads us to speculate that the motility defect exhibited upon Src activation could result from upregulated adhesion with the substrate.

\section{Perspectives}
\noindent Our analysis shows that circular patterns can be used as a tool for estimating hydrodynamic parameters of confined cell monolayers from their mechanical behavior only. In our case, we used this approach to study how hydrodynamic parameters change when the expression level of Src kinase is aberrant. More generally, by using cell lines where molecular ingredients can be modified, circular patterns allow us to explore the links between mechanics and these molecular
components. In this sense, our approach may, for instance, allow to study the upstream/downstream signaling of mechanical polarization, which is often challenging to measure experimentally. Indeed, our joint experimental/theoretical approach provides an operational way to estimate the polarity field from experimental data, like traction forces or mechanical stresses, as confirmed by the good agreement obtained between theoretical results and experimental data. \\

\noindent We have shown here that chiral stresses are negligible at the tissue scale, perhaps because the cells used in this study are themselves achiral. This is consistent with the absence of a clear bias in the direction of rotations between clockwise (CW) and counter-clockwise (CCW) directions: 10/17 CW \textit{vs.} 7/17 CCW in the OFF-ON case and 9/19 CW \textit{vs.} 10/19 CCW  for the OFF-OFF case. Other cell lines exhibit higher chirality and display biased alignement when seeded in rings or linear stripes: NIH cells align in a clockwise way while C2C12 are biased counterclockwise \cite{Wan2011}. Measuring the contribution of chiral stress for these cell lines in circular pattern could provide meaningful insights into the implications of cellular chirality in physiology and pathology, which remains an open question. \\

\noindent In this work, we modulated oncogenic signaling by activating a particular oncogene, the Src kinase. We showed that cells in confluent monolayers tend to spread, exert higher traction forces in norm and reinforce their adhesion with the substrate upon Src activation. Even though we do not know how to infer the biomechanical properties of cells in confluent monolayers from those of  isolated cells, it is interesting to note that Src activation also influences the single cell motion. In particular, we observed that  Src-activated cells diffuse more than non-illuminated cells (Fig.~S8). Understanding how collective cell behaviors emerge from single cell properties is an exciting topic left for future studies. Our study highlights the importance of correct dose of Src signalling to coordinate collective cell migration. Other oncogenes could be activated by light, such as Yap \cite{Toh2022}, and EGFR \cite{Farahani2021}: it would be interesting to test the effect of their over-activation on spontaneous collective rotation. Besides, oncogenes are not the only ones to play a central role in tumor development. The inactivation of tumor suppressor genes seems to play as important a role in tumorigenesis as that of oncogenes \cite{Sherr2004}. Another interesting prospect would be to repeat similar experiments by inactivating a tumor suppressor in order to study the effect of this biological disruption on the collective behavior of cells.

\section{Methods}
\subsection{Cell culture.}
Madin-Darby Canine Kidney (MDCK) optoSrc \cite{Moitrier2019b,Kerjouan2021} and optoSrc-vinculin-iRFP cells were maintained in Dulbecco's Modified Eagle's Medium (Gibco) containing 10\% (vol/vol) fetal bovine serum (FBS, Gibco) and 1\% (vol/vol) penicillin-streptomycin (Gibco) at 37$^{\circ}$C, 5\%~CO$_2$ and humidification. Cells were used at a low passage, between passage 2 to 15 after defreezing (frozen cells are at passage 17), and regularly tested for mycoplasma. 

\subsection{Polyacrylamide gel with circular adhesive patterns.}
The protocol to prepare patterned gels (Fig.~S1) is adapted from \cite{Tse2010,Vignaud2014}. First, we air plasma-clean glass coverslips (\#1.5, 30~mm diameter, Warner Instruments) during 30\,s (Harrick plasma cleaner). A 100~$\upmu$L drop of poly(L-lysine)-graft-poly(ethylene glycol) (PLL-g-PEG, Susos, 0.4 mg/ml in HEPES 10~mM) is then sandwiched between the coverslip and a piece of parafilm. After 1\,hr, we attach the PLL-g-PEG-coated side of the coverslip to the chrome side of a photolithography mask (Compugraphics) harbouring the desired patterns (disks of radius $R=100\,\upmu$m) using a drop of NaHCO$_{3}$ (100~mM in water). Mask and coverslip are exposed to deep UV for 5\,~min (UVO cleaner, Jelight Company Inc.): this burns the PLL-g-PEG at transparent positions drawn in the mask. Those specific positions will be later filled with
fibronectin. From here, protect patterned coverslips from light. The coverslip is detached with teflon tweezer and washed with NaHCO$_{3}$. A drop of 25~$\upmu$g/mL fibronectin (Gibco) in NaHCO$_{3}$ is then sandwiched between the patterned glass coverslip and parafilm and incubated for 1\,hr: fibronectin only adsorbs where there is no PLL-g-PEG. In the meantime, we mix a solution of 40\% acrylamide (Bio-Rad) with a solution of 2\% bis-acrylamide (Bio-Rad) in water in order to make a gel of $\sim$\,40\,kPa according to the table established by Tse and Engler \cite{Tse2010}. We add 0.3\% (vol/vol) of fluorescent beads (FluoSpheres $0.2\,\upmu$m 660/680, Life technologies). To start polymerization, APS (ammonium persulfate, 1\%\,vol/vol, Sigma) and TEMED (Tetra Methyl Ethylene Diamine, 1\textperthousand\,vol/vol Bio-Rad) are added to the solution containing the beads and thoroughly mixed. In parallel, \textit{activated} coverslips are prepared as follow: air plasma-cleaning for 10\,min, incubation with 3-aminopropyltrimethoxysilane (2\% vol/vol in isopropanol, Sigma) for 10\,min with stirring, three washing steps with water, 10\,min stirring in water, incubation with glutaraldehyde (0.5\% vol/vol in water, Sigma) for 30\,min with stirring, three other washing steps with water, 10\,min stirring in water, and finally drying with compressed air. The polymerisation mix is then sandwiched between a patterned coverslip and an \textit{activated} coverslip: this results in the transfer of the fibronectin islands on top of soft gels. When the polymerization is complete (typically 20-30\,min), the sandwiched gel is immersed in PBS, and the coverslip bearing the gel of Young modulus $\sim$\,40\,kPa is carefully detached using a razor blade. These patterned polyacrylamide (PAA) gels can be stored in PBS in the fridge for several days.\\ 

\noindent To use in experiments, the patterned PAA gel is placed in a POC mini-2 (Pecon GmbH, Germany) and first incubated in culture medium for $\sim$\,1\,hr. Then, about $2\cdot10^5$ cells are deposited. After 6-8~h of incubation at 37$^{\circ}$C, 5\% CO2 and 95\% humidity, the culture medium is changed in order to remove unattached cells floating in solution. The gel is left to incubate overnight to have nearly confluent micropatterns the next day. The medium is changed again with fresh culture medium, the POCmini placed under the microscope and the experiment is started. At the end of the acquisition, the cells are detached from the gel by replacing the culture medium with trypsin (Gibco) directly under the microscope to avoid any displacement of the sample. 

\subsection{Live-cell imaging.}
We used a POC mini-2 cell cultivation system for live-imaging. Time-lapse acquisitions were done on an inverted AxioObserver~7 (Zeiss, Germany) equipped with a temperature, CO$_2$ and humidity regulation (Life Imaging Services, Switzerland), a motorized stage (Marzhauser, Germany) and a retiga R6 (QImaging, Canada). A 40x/0.75/Ph2 objective (Olympus, Japan) was used and phase-contrast images were taken for typically 24\,hrs with a time interval between 6 and 15\,min with a 2$\times$2 binning. An image of the fluorescent beads was also taken at each time-point for the traction force estimation. A reference image of the beads in the gel at rest was taken after trypsinization. Phase-light sources was a pE-100 and fluorescent source (far-red excitation) was a pE-300-white (both from CoolLed, United Kingdom). Image acquisition was computer controlled with the software Metamorph (Molecular Devices, USA). Numbers of colonies (patterns) analyzed per conditions are: 17 for the OFF-ON case and 19 for the control situation (OFF only), all of them with a radius of $R=100\,\upmu$m.

\subsection{Blue-light activation.}
Photoactivation of the MDCK optoSrc cells was performed using a Cooled pE-300 and a 'blue' excitation filter (BP 450-490 nm). The light power was measured after the objective with a PM30 optical power-meter (PR100 Thorlabs, US) to be 0.2\,mW. The illumination consisted of a 200-ms pulse of blue light every 3\,min. Given that the half-life time of the complex CRY2/CIBN is 6$\pm$1\,min \cite{Idevall2012}, this ensures an almost continuous activation of the Src oncoprotein. During OFF-ON experiments, monolayers of MDCK were subjected to 3\,hrs OFF then to 3\,hrs ON, see Fig.~\ref{fig_setup}. 

\subsection{Focal adhesions analysis.}
The stable cell line MDCK co-expressing optoSrc-mCherry and vinculin-iRFP was generated by co-lentiviral transduction and cell sorting based on the co-expression of both exogenous fusion proteins. OptoSrc (gift from Olivier Destaing, IAB, Grenoble) and vinculin-iRFP (gift from Mathieu Coppey, PCC laboratory, Paris) were respectively inserted in the pLenti-MCS\cite{Kerjouan2021} and pCDH-CMV-MCS-EF1 lentivectors (Systembio, \#CD510B-1). Since the fluorescent beads used in TFM experiments are labelled in deep red, we made bead-free polyacrylamide gels to image focal adhesions. To ensure that we were under the same conditions, in terms of blue-light power for Src activation, we used our epifluorescence microscope and the same photo-activation protocol, except for the duration of blue illumination, which we limited to 1\,hour. We imaged vinculin before and after this 1-hour photoactivation using a 60x/1.35 oil objective (Olympus, UPlanSAPO) and a long time exposure ($\sim$\,s) to maximize signal collection. To compare the average intensity of vinculin-iRFP between OFF and ON, we computed the mean intensity inside the circular pattern normalized by the mean intensity of a region outside the pattern (background) for $N=21$ colonies imaged (Fig.~S7a). Vinculin-iRFP being not homogeneously expressed within the cells, and so, within the pattern, we selected 15 regions having a roughly homogeneous signal in vinculin-iRFP (Fig. S7b). We then used a pipeline inspired by \cite{Horzum2014} with fewer steps to compare the number of focal adhesions between OFF and ON inside these regions. We first performed a subtract background (sliding paraboloid, rolling ball radius set to 16~pixels), followed by an enhancement of the local contrast (CLAHE, block size = 19, histogram bins = 256, maximum slope = 6, no mask and fast), then the threshold was manually adjusted (same threshold between OFF and ON but different from region to region), and finally, we executed a particle analysis (size above 0.25\,$\upmu$m$^{2}$). These imaging processing steps allowed us to quantify the number of focal adhesion (Fig. S7c).

\subsection{Mean square displacement computation.}
The 2d trajectories of individual cells correspond to the position of the centroid of an ellipse fitted on manually-drawn cell outlines, based on phase-contrast images taken at regular time intervals $\Delta t=3$\,minutes during 1\,hour. The Mean Square Displacement (MSD) at a given time lag $\tau=k\Delta t$ was calculated using the following equation:
$$\mathrm{MSD}(\tau)= \frac{1}{n- k}\sum_{i=1}^{n-k} \left[ (x(t_{i+k}) - x(t_i))^2 + (y(t_{i+k}) - y(t_i))^2 \right]\,,$$
where $(x(t), y(t))$ is the position of the centroid at time $t$ and $n$ is the total number of time points. MSD values were then averaged over $N=45$ independent trajectories. The results were plotted as MSD versus $\tau$ (Fig.~S8). Since the statistical errors related to MSD computation become large as $\tau$ increases, MSD curves were plotted only for the first quarter of all time points.

\subsection{Mechanical state of the confined monolayer.}
For the control experiments (OFF-OFF case), we defined for each pattern, the time-point for which the confined monolayer is confluent, as the reference time $t=0$. This allowed us to synchronise the different movies. We then limited the duration of analysis to 6\,hrs by temporally cropping all the different movies from $t=0$. This allowed to ensemble-average all the movies corresponding to the control situation. For the OFF-ON experiments, we could not easily synchronize the movies, so we considered only the patterns for which the confluence was reached at the start of the acquisition, and ensemble-averaging was performed without synchronization. To estimate the time-evolution of the cell density, the number of cells of each colony was manually counted using phase-contrast images at three relevant time-points : 0\,, 3\, and 6\,hour (Fig.~S4a-b). For $t=0$, the cell density was estimated to be $(1.9 \pm 0.5)\cdot10^5$\,cells/cm$^2$ (standard deviation, $N=19$) for the control case and $(2.2 \pm 0.7)\cdot10^5$\,cells/cm$^2$ (standard deviation, $N=17$) for the OFF/ON case. Since we only consider data inside the adhesive pattern ($r\leq R$), all fields (velocity, traction forces and stresses) were masked using a binary image of the inital pattern of radius $R$ after PIV/TFM computations.\\

\noindent All the ImageJ plugins we refer to in the following sections have been developed by Tseng \textit{et al.} and described in \cite{Martiel2015}.

\paragraph*{Alignment.}
First, the Template Matching plugin is used with bead images in order to align the different frames with the reference image (taken after trypsinization). Then, phase-contrast images are corrected for this drift.

\paragraph*{Velocity field.}
Velocity fields in the cell monolayer were measured with particle image velocimetry (PIV) using the MatPIV toolbox \cite{Sveen2004} on the aligned phase-contrast images. The interrogation window size was set to 128\,pixels ($\sim\,$30.7 \,$\upmu\mathrm{m}$), with an overlap of 50\%. Filters were used to remove the outliers: a global filter that excludes velocities greater than 3 times the standard deviation of the velocity field and a local filter that eliminates velocities that are substantially different from their neighbors (2 times higher than the median of the neighboring speeds). A binary mask corresponding to the location of the circular pattern was used to filter data outside the pattern. Velocities were then linearly interpolated using Matlab built-in \texttt{interp2.m} function on the TFM grid (vector spacing: 16\,pixels). We then used a cartesian-to-polar transformation to obtain the velocity field in polar coordinates: $u_r(r,\theta,t),\, u_{\theta}(r,\theta,t)$ expressed in $\upmu$m.h$^{-1}$. 

\paragraph*{Traction forces.}
Traction forces were computed with ImageJ as follows. After an alignment step, the substrate displacement field was estimated on the bead images using the Iterative PIV plugin in advanced mode with the following values for interrogation window/search window/vector spacing for 3 successive passes : 128/256/64 for the first, then 64/128/32 and finally 32/64/16 (pixels). Traction forces were computed with the FTTC plugin (regularization factor set to 8$\cdot10^{-11}$, Poisson ratio to 0.5 and Young modulus to 40\,kPa). Finally, we also used a cartesian-to-polar transformation in Matlab to obtain traction forces in polar coordinates: $T_r(r,\theta,t),\, T_{\theta}(r,\theta,t)$, expressed in Pa. The quality of force reconstruction was assessed by computing the  quality factor, $Q$, which compares the vector sum of the traction forces of the monolayer with the total force exerted by the monolayer on its substrate (sum of the norm of the forces):
$$ Q= 1 - \frac{||\sum \vec{T}||}{\sum ||\vec{T}||} $$
Force reconstruction was considered acceptable for $Q>0.85$ for all the time-points of the 6\,hrs period (\cite{Maruthamuthu2011}, see~Fig.~S9).

\paragraph*{Stress inference.}
\label{sec:stress}

The mechanical stress tensor $\sigma$ was estimated by
Bayesian Inversion Stress Microscopy (BISM, see \cite{Nier2016}
for additional details) from the traction force vector field
$\vec{T}$ obtained by TFM. For each image, the dimensionless regularization
parameter was typically of the order of $\Lambda = 10^{-5}$, as determined using
the L-curve method. In practice, we performed Bayesian inversion
on the smallest square domain containing the cell confinement disk,
using Cartesian coordinates. This approach was validated numerically
as follows (Fig.~S10). As in \cite{Nier2016},
we used the finite element software \texttt{FreeFem++} \cite{Hecht2012} 
to solve numerically, in a circular domain, the force balance equations
for a viscous material driven by active force dipoles and thus generated
stress and traction force numerical data. Both sets, obtained on a disk, were
complemented by zero values to fill a square domain of lateral extension
identical to the disk diameter. The BISM algorithm was next applied
to the traction force data set, using a zero-stress  boundary condition
on edges of the square domain (\textit{i.e.} $\sigma_{ij} \, n_j=0$  
where $\vec{n}$ denotes the vector normal to the edge, and 
summation over repeated indices is implied).
As a measure of accuracy, we computed the coefficient of determination,
$R^2$ that compares a numerical data set $\{\sigma^{\mathrm{num}}\}$ of stresses, 
to the set $\{\sigma^{\mathrm{inf}}\}$ of inferred stresses, and typically found
excellent agreement between the two data sets ($R^2 > 0.9$).
An example of our numerical results is given in Fig.~S10.
When applying the same protocol to experimental TFM data sets,
we checked that the relationship relating the spatially-averaged
monolayer tension (or isotropic stress $\sigma_{\mathrm{iso}}$)
to the first moment of the radial traction force component  ($ \langle \sigma_{\mathrm{iso}} \rangle = - \frac{1}{2}\langle T_r r \rangle$, see \cite{Landau1975})
was verified for the inferred tension, further validating our stress
estimation.


\section*{Author contributions}
N.P. and I.B. designed the research. N.P., F.M. and I.B. performed the experiments. P.M. and C.B.-M. developed the theoretical model. All the authors contributed to analysing the experimental data and writing the manuscript. 

\section*{Conflicts of interest}
There are no conflicts of interest to declare.

\section*{Data availability}
All data analyzed in this study are available in a public Zenodo repository:\\
doi.org/10.5281/zenodo.12667174


\section*{Appendix}
\subsection*{Appendix A: Role of achiral and chiral active stresses.}
\label{app:A}

\setcounter{figure}{0}
\renewcommand{\figurename}{Fig.}
\renewcommand{\thefigure}{A\arabic{figure}}

\begin{figure}[h!]
\centering
	\includegraphics[width=\linewidth]{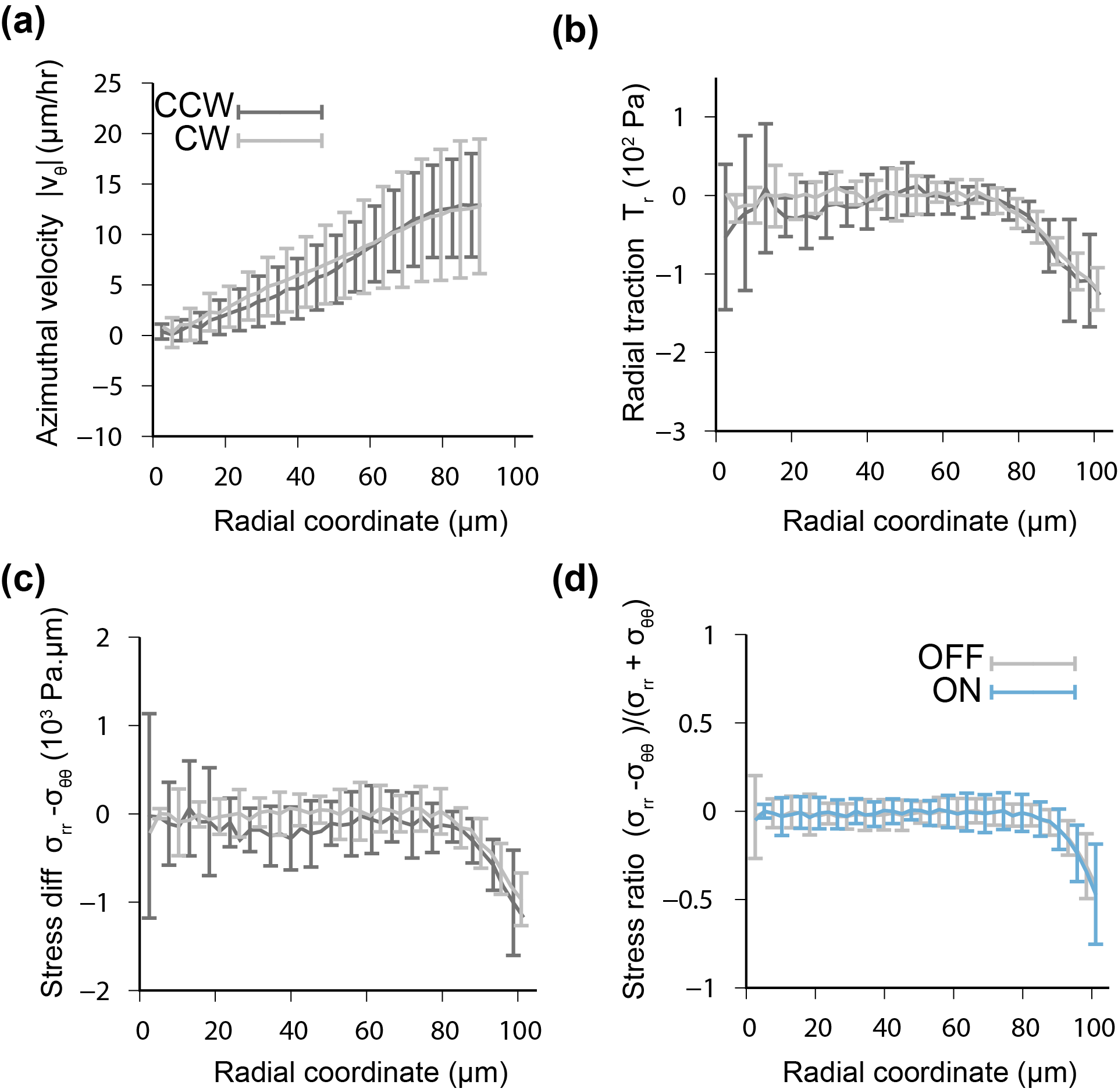}
\caption{Chiral and achiral active stresses are negligible. Comparison of radial profiles between cell monolayers that rotated in the clockwise direction (light  gray) and that rotated in the counter-clockwise direction (darkgray) for the first 3\,hrs of the OFF-ON case. (a) Azimuthal component of cell averaged velocity field. (b) Radial component of cell averaged traction force field. (c) Stress difference $\sigma_{rr}-\sigma_{\theta\theta}$. Plots correspond to data averaged over $N=10$ colonies with clockwise rotation and $N=7$ colonies with counter-clockwise rotation. (d) Radial profile of the stress ratio $(\sigma_{rr}-\sigma_{\theta\theta})/(\sigma_{rr}+\sigma_{\theta\theta})$.  Light gray: 3\,hrs OFF, and light blue: 3\,hrs ON. Error bars correspond to standard deviation after azimuthal and temporal averaging.}
\label{figA1}
\end{figure}

To estimate the importance of active stresses in driving collective rotation of cell monolayers, we performed the following analyses. The theoretical framework introduced in Section~\ref{sec:3} was generalised to include achiral and chiral active stresses with amplitudes $\zeta$ and $\tau$, respectively. These active processes are sustained by consuming a chemical fuel, typically adenosine triphosphate hydrolysis in the context of cell cultures, and the chemical potential difference of this reaction is denoted as $\Delta\mu$. As explained in the Electronic Supplementary Information and in Refs.~\cite{BlanchMercader2021b,Hoffmann2020chiral}, the general expression of the total stress components \eqref{eq:StressTensorRR} and \eqref{eq:devStressTensorRTheta} reads 
\begin{align}
\sigma_{rr,\theta\theta}&=-P
\pm\frac{\kappa}{2}\left(\frac{S^2}{r^2}-(S')^2\right) \mp\frac{\zeta\Delta\mu}{2}S^2 \cos(2\psi)\mp\frac{\tau\Delta\mu}{2}S^2 \sin(2\psi)\label{eq:StressTensorRRApp}  \\
\sigma_{r\theta,\theta r}&=\eta\left(\partial_r v_{\theta}(r)-\frac{v_\theta(r)}{r}\right)-\frac{\zeta\Delta\mu}{2}S^2 \sin(2\psi)+\frac{\tau\Delta\mu}{2}S^2 \cos(2\psi)\label{eq:devStressTensorRThetaApp} 
\end{align}

\noindent where the upper (lower) signs correspond to the first (second) index pair. Note that the trace of the total stress tensor is $\sigma_{rr}+\sigma_{\theta\theta}=-2P$. In experiments, $10/17$ of the cell monolayers exhibit collective rotation in the clockwise direction and $7/17$ in the counter-clockwise direction. The amplitude of the averaged azimuthal velocity for these two directions are similar (Fig.~\ref{figA1}a). Changing between the clockwise to the counterclockwise direction in our continuum description corresponds to applying the transformation $v_\theta\rightarrow -v_\theta$. Because the driving processes depend only on the polarity field and its gradients, this transformation can be mapped into a re-orientation of the polarity field. Specifically, in the traction force or achiral active stress dominated regimes, the polarity orientation transforms as $\psi\rightarrow -\psi$, and in the chiral active stress dominated regime, it is $\psi\rightarrow \psi+\pi/2$. These transformations change the sign of the corresponding active process in \eqref{eq:devStressTensorRThetaApp} and \eqref{eq:TractionO}, and therefore reverse the direction of rotation. This suggests that the polarity of cells varies between the two rotating states. \\

\noindent To investigate the re-orientation of the cell polarity between the two rotation modes, we compared their steady-state stress and traction force patterns. In particular, we focus on two variables:  the radial component of the traction force field $T_r$ and the stress difference $\sigma_{rr}-\sigma_{\theta\theta}$. In our theoretical framework, the expression for these quantities is
\begin{align}
T_r&=-T_0 S(r)\cos(\psi)\\
\sigma_{rr}-\sigma_{\theta\theta}&=-\zeta\Delta\mu S^2 \cos(2\psi)-\tau\Delta\mu S^2\sin(2\psi)+\kappa\left(\frac{S^2}{r^2}-(S')^2\right) \label{eq:stressdiff}
\end{align}
and their dependence on the polarity field is explicit through $S$ and $\psi$. Depending on the dominant active process for collective rotation, these two variables may be different between clockwise and counterclockwise rotating states. In experiments, we found no significant differences (Fig.~\ref{figA1}b-c). Therefore, we conclude that chiral active stress is negligible in cell monolayers confined to discs.  \\

\noindent To further estimate the relative importance between traction forces and achiral active stress, we analysed the ratio between the stress difference $\sigma_{rr}-\sigma_{\theta\theta}$ and the trace of the stress $\sigma_{rr}+\sigma_{\theta\theta}$ (Fig.~\ref{figA1}d). Except for a region near the boundary, this ratio is much smaller than 1, suggesting that $\sigma_{rr}\sim\sigma_{\theta\theta}\sim-P$, see Eq.~\eqref{eq:StressTensorRRApp}. Therefore the achiral active stresses are subdominant in comparison to traction forces.\\

\noindent From the stress difference $\sigma_{rr}-\sigma_{\theta\theta}$ at the confinement interface in Fig.~\ref{figA1}d, assuming that $\tau \Delta \mu = 0$, and taking the parameters from Table~\ref{tab:table1}, we estimated the order of magnitude of $\zeta\Delta\mu+\chi \sim 10^3$~Pa$\cdot\upmu$m based on Eq.~\eqref{eq:stressdiff}. When $\chi\ll\zeta\Delta\mu$, one obtains that the ratio between the amplitudes of the two active processes $\zeta\Delta\mu/R T_0\sim0.05$ for $\zeta\Delta\mu\sim 10^3$ Pa$\cdot\upmu$m. When $\chi\gg\zeta\Delta\mu$, then achiral active stresses are negligible and in this case $\chi\sim 10^3$~Pa$\cdot\upmu$m, and $\kappa=L_c^2\chi\sim 10^5$~Pa$\cdot\upmu$m. Altogether, our analysis leads us to conclude that traction forces are the dominant force generation mechanism.

\subsection*{Appendix B: Theoretical fits for cell monolayers in the OFF-OFF case.}
\label{app:B}

\setcounter{figure}{0}
\renewcommand{\figurename}{Fig.}
\renewcommand{\thefigure}{B\arabic{figure}}

\begin{figure}[!ht]
\centering
	\includegraphics[width=\linewidth]{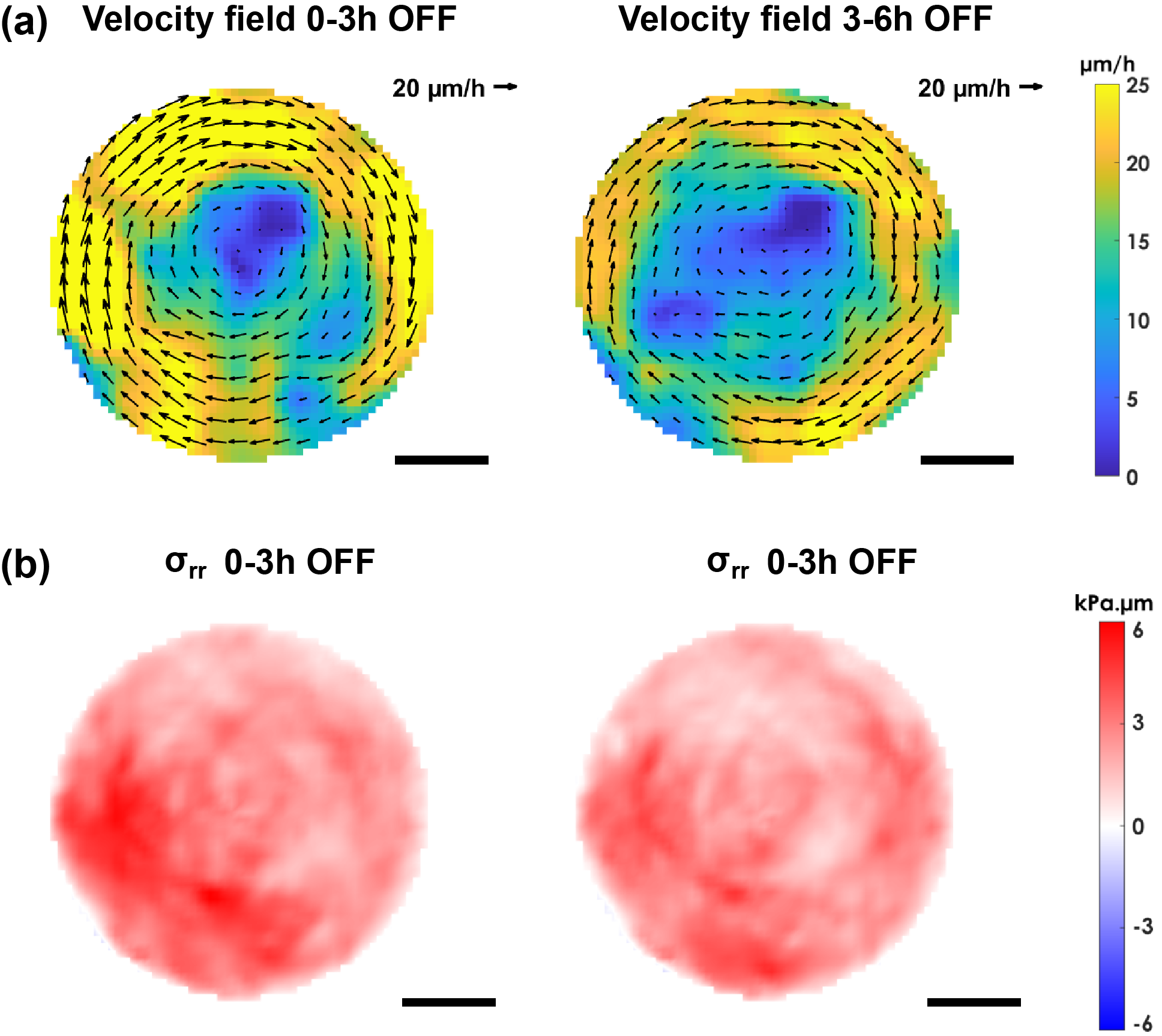}
\caption{Velocity and normal stress radial component during 6\,hrs OFF. (a) Time-averaged velocity magnitude contour plot superimposed with velocity vector plot obtained by PIV for the first 3\,hrs (OFF, left) and for the last 3\,hrs (OFF, right) for a given colony. Only one arrow out of three is drawn for visibility. (b) Time-averaged stress component $\sigma_{rr}$ measured by BISM over the first 3\,hrs (OFF, left) and over the last 3\,hrs (OFF, right), same colony as in a. Scale bars: 50~$\upmu$m. Note, that, in 2d, stresses are expressed in kPa$\cdot\upmu$m. See Movie~1 for the spatio-temporal dynamics of the velocity field and the stress field.} 
\label{fig_B1}
\end{figure}

\begin{figure}
\centering
	\includegraphics[width=\linewidth]{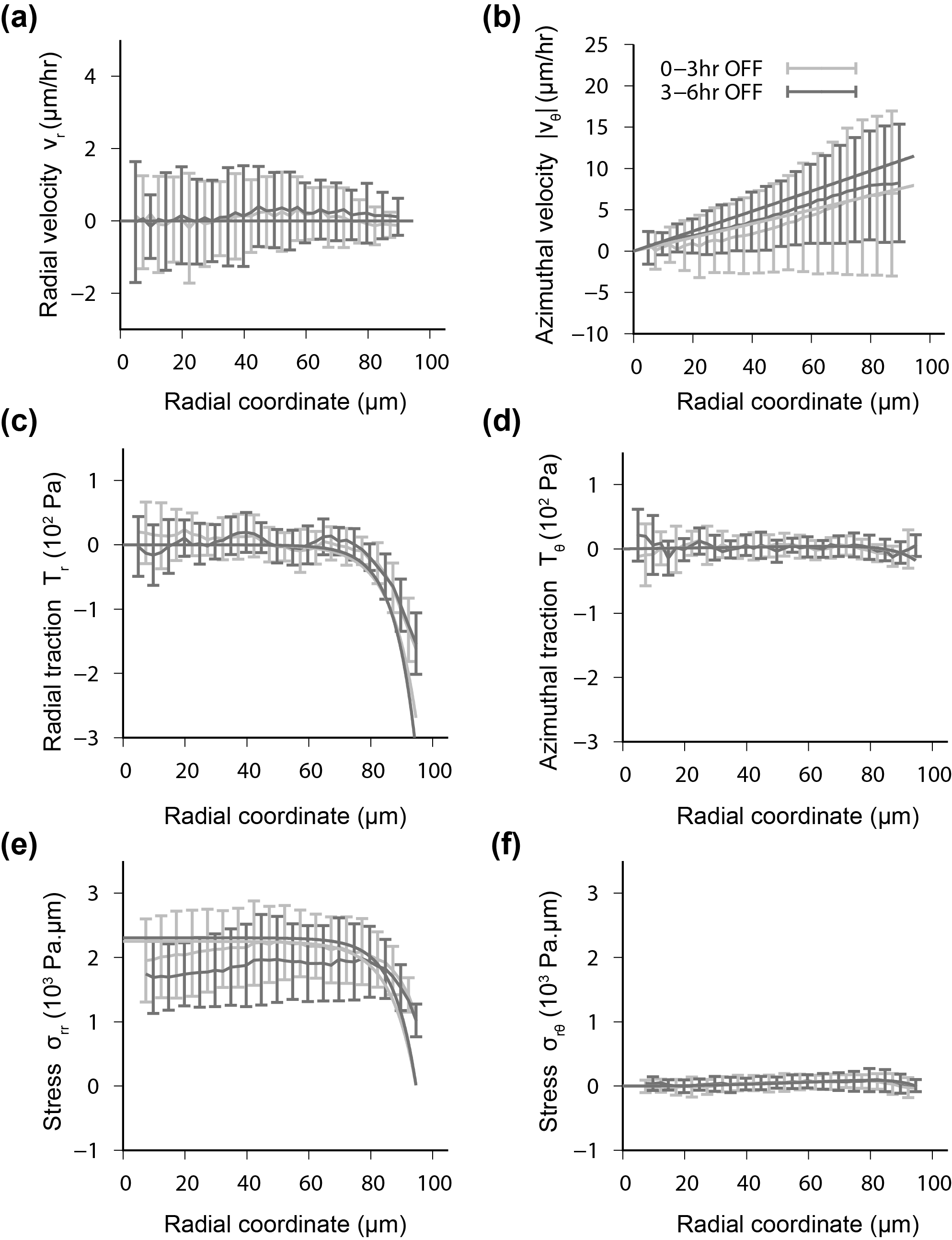}
\caption{Radial experimental profiles in the 0-3h (light gray) and 3-6h (dark gray) of the OFF-OFF case with theoretical fits. (a-b) Radial profiles of the radial and azimuthal components of cell averaged velocity field, respectively. (c-d) Radial profiles of the radial and azimuthal components of cell averaged traction force field, respectively. (e-f) Radial profile of the components of the stress $\sigma_{rr}$ and $\sigma_{r\theta}$, respectively. Plots correspond to data averaged over $N=19$ colonies and error bars correspond to standard deviation after temporal averaging. Theoretical fits are shown as solid curves in light gray for the 0-3h OFF case and dark gray for the 3-6h OFF case. Fitted parameter values are in Table~\ref{tab:table1}.} 
\label{fig_B2}
\end{figure}

\noindent In this appendix, we analyse the collective rotation of cell monolayer with a 6\,hrs OFF period (OFF-OFF case), and therefore in the absence of Src 
activation. Fig.~\ref{fig_B1} shows representative velocity and stress maps of rotating cell monolayers. We performed the same fitting procedure as explained in Section~\ref{sec:3.2} and obtained the results shown in Fig.~\ref{fig_B2}. In the absence of Src 
activation, we found that the averaged experimental profiles and the corresponding theoretical fits are similar between the first 3\,hrs and the next 3. The values of the fitted parameters can be found in Table.~\ref{tab:table1}. No difference was found on the fitted parameter values between these two time periods.

\FloatBarrier
\section*{Acknowledgements}
The authors would like to thank O. Destaing, C. Duclut and K. Guevorkian for their constructive comments on the manuscript, their team members for discussions, as well as Fanny Cayrac from the BMBC facility of PCC laboratory for the Vinculin-iRFP plasmid insertion. This work was supported by Les Entreprises Contre le Cancer Paris Ile-de-France and received additional support from grants ANR-11-LABX-0038, ANR-10-IDEX-0001-02. NP was supported by a doctoral fellowship from the IPGG and from the Fondation ARC pour la recherche sur le cancer (grant ARCDOC42021020003340).


\bibliographystyle{unsrt} 
\bibliography{main} 

\newpage
\FloatBarrier

\begin{center}
  {\Large Electronic Supplementary Information}
\end{center}

\setcounter{section}{0}

\section{Theoretical model}
In this section, we detail the theoretical framework used to describe monolayers of cells \cite{BlanchMercader2021b}. For more details on the active gel theory we refer to Refs~\cite{kruse2005generic,joanny2007hydrodynamic,julicher2018hydrodynamic}.\\

\noindent The cell monolayers are characterised by the coarse-grained cell velocity field $\mathbf{v}$ and the coarse-grained cell polarity field $\mathbf{p}$. As explained in the main text, we consider the system incompressible and as a consequence the cell number density $n$ is constant. Because source and sinks of mass, such as cell division or apoptosis, are ignored, the incompressibility condition reduces to $\nabla.\mathbf{v}=0$. 
\noindent In the absence of inertial effects, the momentum conservation equation reduces to the force balance equation
\begin{align}
\partial_\beta \sigma_{\alpha\beta}&=T_\alpha\label{eq:ForcBalance}
\end{align}
The gradients of total 2d stresses $\sigma$ are balanced by the traction forces $\mathbf{T}$ that arise from interaction forces between cells and the underlying substrate. For the traction forces, we consider the special case
\begin{align}
T_\alpha=\xi v_\alpha-T_0 p_\alpha\label{eq:TractionFor}
\end{align}
that includes two contributions that is a viscous friction proportional to the velocity field $\mathbf{v}$, with a friction coefficient $\xi$, and active traction forces proportional to the polarity field $\mathbf{p}$, with an amplitude $T_0$. For a more general discussion of the traction forces, we refer to \cite{julicher2009generic}.
\\

\noindent To build the constitutive relations of the system, we follow the standard approach of non-equilibrium thermodynamics~\cite{de2013non}. It consists of first identifying pairs of conjugated thermodynamic forces and fluxes by inspecting the time derivative of the free energy. In a second step, the fluxes are expressed to linear order in terms of the forces, where the coupling coefficients obey the Onsager relations. \\

\noindent Here, we choose the following quantities as thermodynamic forces~\cite{kruse2005generic,joanny2007hydrodynamic,julicher2018hydrodynamic}: the symmetric part of the velocity gradient tensor with components $v_{\alpha\beta}=(\partial_\alpha v_\beta+\partial_\beta v_\alpha)/2$, the field $\mathbf{h}=-\delta\mathcal{F}/\delta\mathbf{p}$, where  $\mathcal{F}$ is the equilibrium free energy, and the chemical potential difference $\Delta\mu$ of the fuel consumption reaction, typically ATP hydrolysis in cells. The corresponding thermodynamic fluxes are given by the deviatory stress tensor $\mathsf{\sigma}^{d}$, the co-rotational convective derivative of the polarization field $D\mathbf{p}/Dt$, and the rate $r$ of ATP-hydrolysis. The deviatory stress tensor is defined as $\mathsf{\sigma}^d=\mathsf{\sigma}-\mathsf{\sigma}^e$ the difference between the total stress tensor $\mathsf{\sigma}$ and the Ericksen stress $\mathsf{\sigma}^e$. The co-rotational convective derivative of the polarization field is given by 
\begin{align}
\frac{D p_\alpha}{Dt} &=\partial_t p_{\alpha}+v_\beta \partial_\beta p_\alpha+\omega_{\alpha\beta}p_\beta.\label{eq:corotCovecDerivative}
\end{align}
where $\omega_{\alpha\beta}=(\partial_\alpha v_\beta-\partial_\beta v_\alpha)/2$ is the antisymmetric part of the velocity gradient tensor. As we assume $\Delta\mu$ to be constant, we do not consider $r$ any further.

\noindent The deviatory stress and the co-rotational convective derivative of the polarization field are
\begin{align}
\sigma_{\alpha\beta}^d& = 2\eta v_{\alpha\beta}+\frac{\nu}{2}\left(p_{\alpha}h_\beta+p_\beta h_{\alpha}-p_\gamma h_\gamma \delta_{\alpha\beta}\right)\nonumber \\&-\zeta\Delta\mu\left(p_{\alpha}p_\beta-\frac{1}{2}p_\gamma p_\gamma\delta_{\alpha\beta}\right)+\frac{1}{2}\tau \Delta \mu(p_\alpha p^\perp_\beta+p_\beta p_\alpha^\perp)\label{eq:devstresstensor}\\ 
\frac{D p_\alpha}{Dt}&= \frac{h_\alpha}{\gamma}-\nu v_{\alpha\beta}p_\beta\label{eq:dinamicadirector}
\end{align}
where we used the incompressibility condition $v_{\gamma\gamma}=0$ to simplify these expressions, and redefine the pressure field $P$ such that $\sigma_{\gamma\gamma}^d=0$. In Eq.~\eqref{eq:devstresstensor}, the first  term accounts for viscous stresses, where the coefficient $\eta$ is the shear viscosity. The second term couples the mechanical stresses to the field $\mathbf{h}$. The remaining terms couple the mechanical stress to ATP-hydrolysis. The term proportional to the coefficient $\zeta$ denotes the achiral  components of the active stresses, whereas the term proportional to the coefficient $\tau$ denotes the chiral components of the active stresses \cite{Hoffmann2020chiral}. The field $\mathbf{p}^\perp$ corresponds to the perpendicular vector to the polarity field. In Eq.~\eqref{eq:dinamicadirector}, the first term captures relaxation of the polarization field with $\gamma$ being a rotational viscosity. The second term describe the re-orientation of the polarity field to shear flows. The coefficient $\nu$ is the so-called shear flow-alignment parameter. Note that, in this equation, we have omitted an active term, that is a coupling to $\Delta\mu$. One of these terms would be of the form $\lambda_1\Delta\mu p_\alpha$ and it amounts to a renormalization of parameters \cite{kruse2005generic,joanny2007hydrodynamic}. Another of such terms would be of the form $\lambda_2\Delta\mu p_\alpha^\perp$, which describes the active spinning of the polarity field, and it necessitates a substrate to exists.  \\ 

\noindent Explicit expressions for the Ericksen stress $\mathsf{\sigma}^e$ and the field $\mathbf{h}$ are obtained by fixing the equilibrium free energy  $\mathcal{F}$ of the system. We choose:
\begin{align}
\mathcal{F}&=\int_\mathcal{A}\left\{\frac{\chi}{2}p_{\alpha}^2+\frac{\kappa}{2}(\partial_{\alpha}p_{\beta})^2\right\}da. \label{eq:freeenergy} 
\end{align}
The two terms capture the elastic energy associated with distortions of the polarization field similar to the free energy used for liquid crystals~\cite{de1993physics}. Here, we consider $\chi>0$ meaning that the preferred bulk equilibrium state is disordered. The energy cost associated with gradients of the polarization field is accounted for by the second term. It is equal to the Frank energy in the one-constant approximation with modulus $\kappa$.

\noindent Based on the free energy \eqref{eq:freeenergy}, the expression for the Ericksen stress is  \begin{align}
\mathsf{\sigma}^e_{\alpha\beta}&=-P\delta_{\alpha\beta}-\kappa\left(\partial_\alpha p_\gamma \partial_\beta p_\gamma-\frac{1}{2}\partial_\rho p_\gamma \partial_\rho p_\gamma\delta_{\alpha\beta}\right)+\frac{1}{2}\left(p_\alpha h_\beta-p_\beta h_\alpha\right) \label{eq:EricksenS} 
\end{align}
and of the molecular field $\mathbf{h}$ is
\begin{align}
h_\alpha&=-\frac{\delta {\cal F}}{\delta p_\alpha}=-\chi p_{\alpha}+\kappa\partial_{\gamma\gamma}p_{\alpha}. \label{eq:molfield} 
\end{align}

\noindent To obtain the continuum equations in the main text, we make the following simplifying approximations. In Eq.~\eqref{eq:dinamicadirector}, we approximate $D p_\alpha/Dt\approx 0$ and ignore the effect of the shear-flow alignment by setting $\nu=0$. Besides, the angle between the polarity field and the radial direction $\psi$ is considered to be uniform. In Eq.~\eqref{eq:devstresstensor}, the stress that arises from couplings with the field $\mathbf{h}$ and both active stresses are ignored by setting $\nu=\zeta=\tau=0$. In addition, the off-diagonal components of the Ericksen stress are ignored. The importance of some of these approximations in describing cell monolayers is discussed in appendix A. 

\section{Additional expression of steady-state mechanical fields}

Below, we provide the analytical solutions of additional mechanical fields
\begin{eqnarray}
T_r(r)&=&\frac{-T_o \cos{(\psi)}I_1(r/L_c)}{I_1(R/L_c)}\\
T_\theta(r)&=&\frac{T_o\sin(\psi) L_\eta}{(L_c^2-L_\eta^2)}\frac{L_\eta I_1(r/L_c)I_2(R/L_\eta)-L_c I_1(r/L_\eta)I_2(R/L_c)}{I_1(R/L_c)I_2(R/L_\eta)}\\
\sigma_{r\theta}(r)&=&\frac{T_o\sin(\psi) L_\eta^2 L_c}{(L_c^2-L_\eta^2)}\frac{ I_2(r/L_c)I_2(R/L_\eta)-I_2(r/L_\eta)I_2(R/L_c)}{I_1(R/L_c)I_2(R/L_\eta)}\\
\sigma_{\theta\theta}(r)&=&-P(r)=T_oL_c\cos(\psi) \frac{I_0(\frac{R}{L_c})-I_0(\frac{r}{L_c})}{I_1(\frac{R}{L_c})}
\end{eqnarray}
which complement Eqs.~(8-10) in the main text. The function $I_n(x)$ is a modified Bessel function of the first kind of order $n$. The parameter $R$ corresponds to the radius of the circular confinement, and $L_c=\sqrt{\kappa/\chi}$ and $L_\eta=\sqrt{\eta/\xi}$ are two characteristic lengthscales of the system. The other parameters are defined above and for more details, see also the main text.

\newpage 
\section{Supplementary Figures}
\setcounter{figure}{0}
\renewcommand{\figurename}{Fig.}
\renewcommand{\thefigure}{S\arabic{figure}}

\begin{figure}[!htb]
\centering
	\includegraphics[width=10cm]{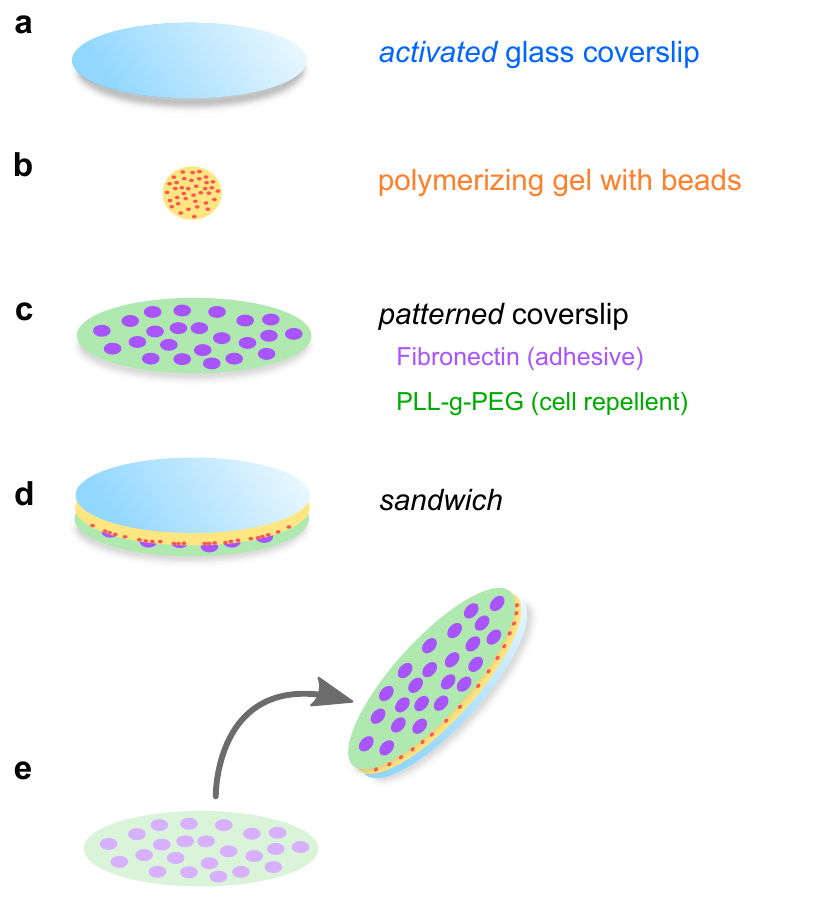}
\caption{\textbf{Preparation of patterned gels for Traction Force Microscopy.} \textbf{a} A silanized glass coverslip is \textit{activated} with glutaraldehyde. \textbf{b} Mix of acrylamide and bis-acrylamide necessary to obtain specific stiffness, containing fluorescent beads. \textbf{c} Circular islands of fibronectin are micropatterned on PLL-g-PEG-coated coverslip using chrome mask and deep UV exposure. \textbf{d} The drop of polymerizing gel is \textit{sandwiched} between the two coverslips. \textbf{e} Once the glass coverslips are detached from each other, the \textit{activated} coverslip will be covered with polyacrylamide gel coated with a layer of cell repellent PLL-g-peg and cell adhesive fibronectin patterns. See Methods for details.}
\label{fig_protocol_gel}
\end{figure}

\begin{figure}[!htb]
\includegraphics[width=1.2\linewidth]{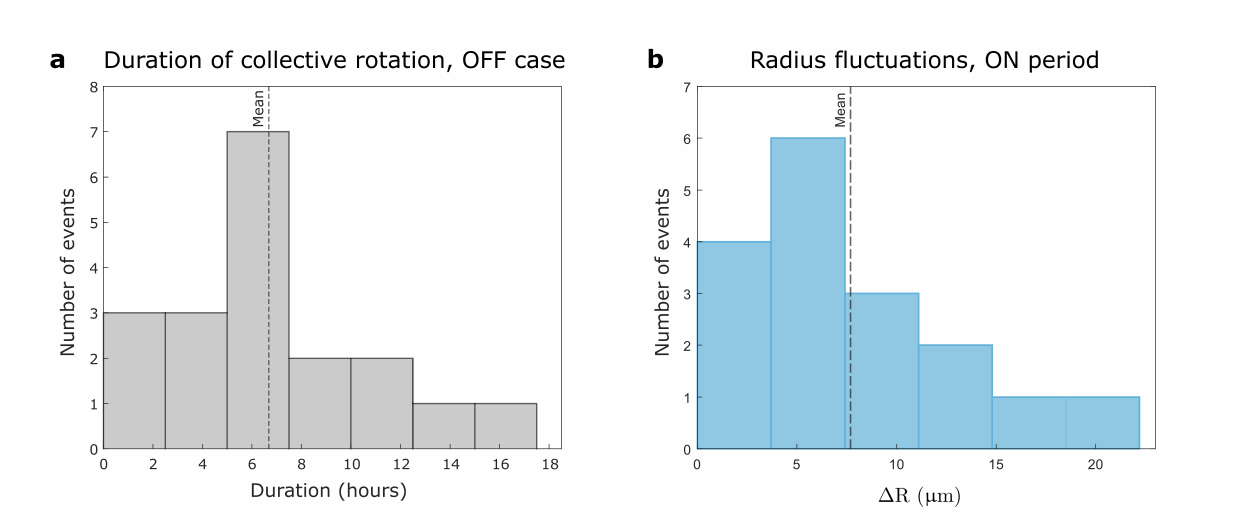}
\caption{\textbf{a-Duration of collective-rotation events for non-activated MDCK optoSrc cells.} Monolayers of MDCK optoSrc cells were confined in 100 $\upmu$m-radius disk and kept indefinitely in the dark. The duration of the collective rotation was measured manually, we found an average rotation time of 7$\pm$4\,hours (standard deviation, $N=19$ colonies) of uninterrupted rotation. \textbf{b-Fluctuations in radius $\Delta R$ during Src activation.} $\Delta R$ was estimated by considering the time point during the ON period (often, but not necessarily, the last) for which the colony area $A$ is maximal: $\Delta R = \frac{R \Delta A}{2A}$ with $\Delta A = A_{\mathrm{max}} -A$ where $A=\pi R^2$ being the area of the circular pattern ($N=$17~colonies).}
\label{fig_period_rotation}
\end{figure}

\begin{figure}[!htb]
\thisfloatpagestyle{empty}
\centering
\includegraphics[width=1.1\textwidth]{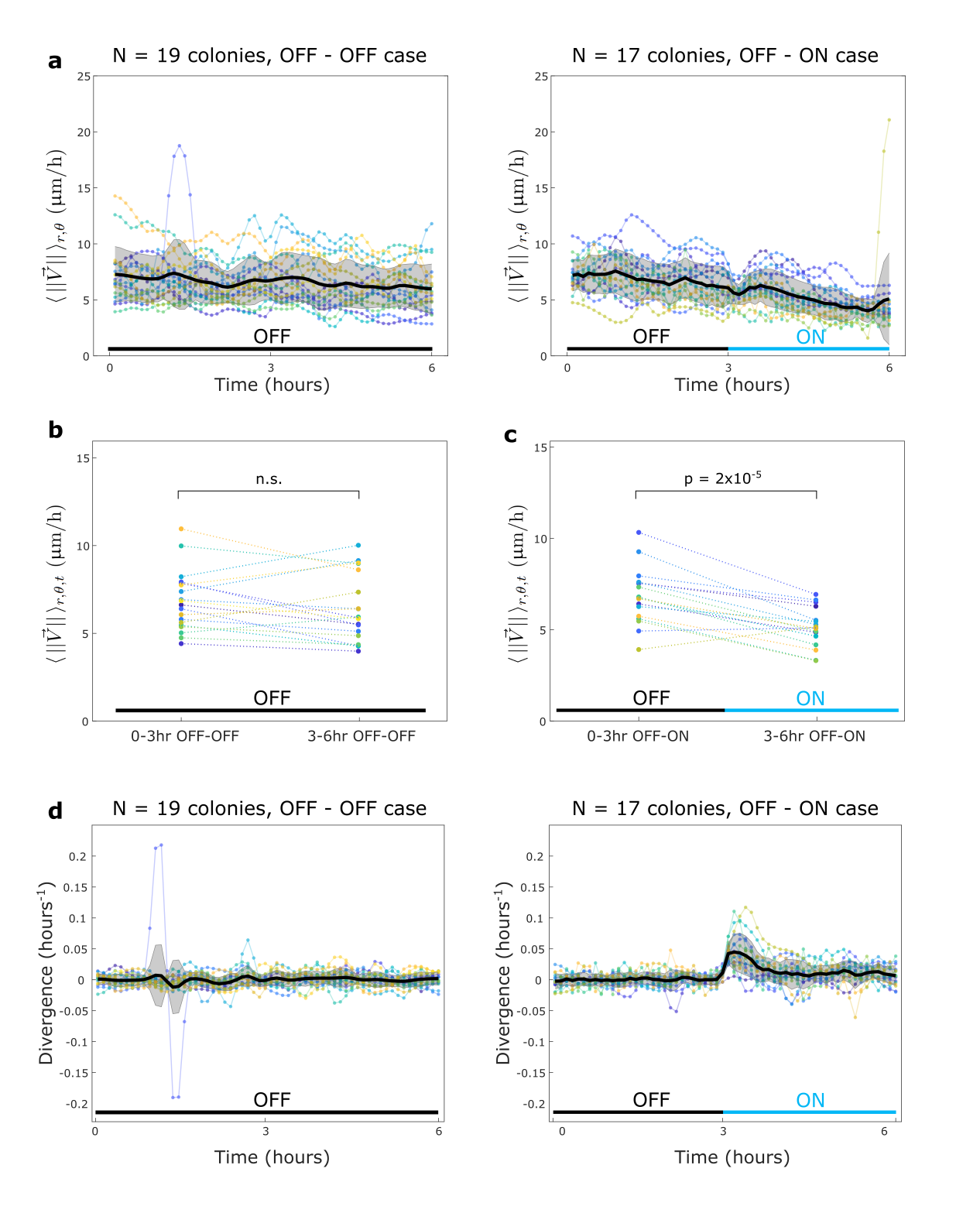}
\caption{\textbf{Src activation reduces velocity magnitude.} \textbf{a} Time evolution of the space-averaged speed $\left( \langle\,||\vec{V}|| \,\rangle_{r,\theta} \right)$ for all colonies (one color = one colony) for control experiments (left) and Src activation experiments (right). \textbf{b-c} For each colony, we time-averaged $\langle\,||\vec{V}|| \,\rangle_{r,\theta}$ over the two periods (first 3\,and last 3\,hrs) for the control case (b) and during Src activation (c). T-tests on paired data were used after verifying normality (Shapiro-Wilk test), n.s. stands for non significant when $p>0.05$. \textbf{d} Spatially-averaged divergence of the velocity field during OFF-OFF (left) and OFF-ON experiments (right). In panels a and d, black lines represent population averages and shaded areas indicate one standard deviation (one color = one colony).}   
\label{new_FigS3_velocity_norm}
\end{figure}

\begin{figure}[!htb]
\centering
\includegraphics[width=1.2\linewidth]{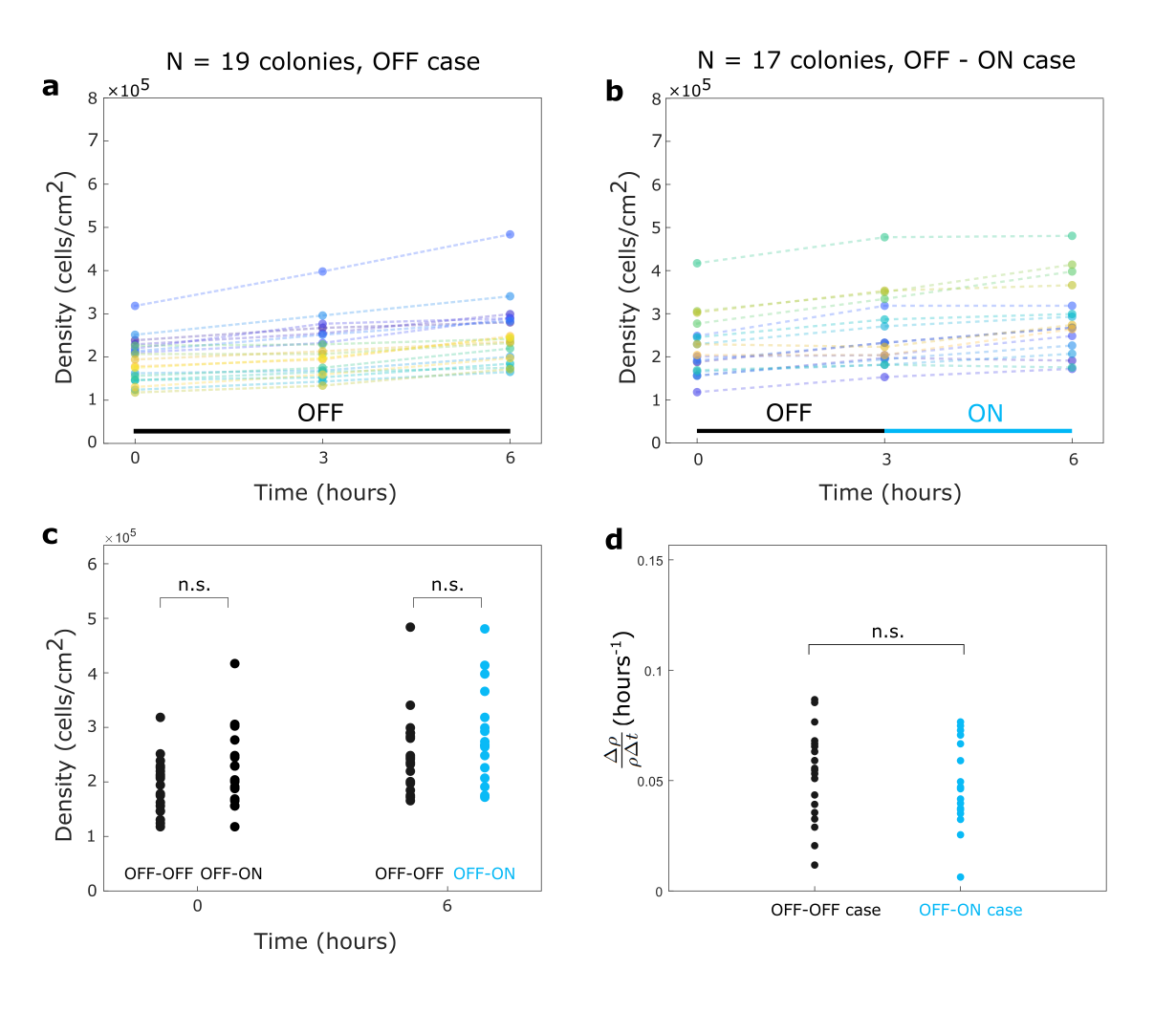}
\caption{\textbf{Time evolution of cell density.} \textbf{a-b} Cell density of all colonies (one color = one colony) at three different time-points (0, 3\,hrs, 6\,hrs) for control experiments (a) and during Src activation (OFF-ON case, b). \textbf{c} Comparison of initial ($t=0$, left) and final ($t = 6$\,hrs, right) cell density between the two types of experiments. \textbf{d} Density variation rate over the course of experiment ($\Delta t=6\,$hrs) for control experiments and during Src activation. Mann Whitney-U tests indicate no significant difference (n.s. for $p>0.05$).}  
\label{new_FigS4_density}
\end{figure}

\begin{figure}[!htb]
\centering
	\includegraphics[width=\linewidth]{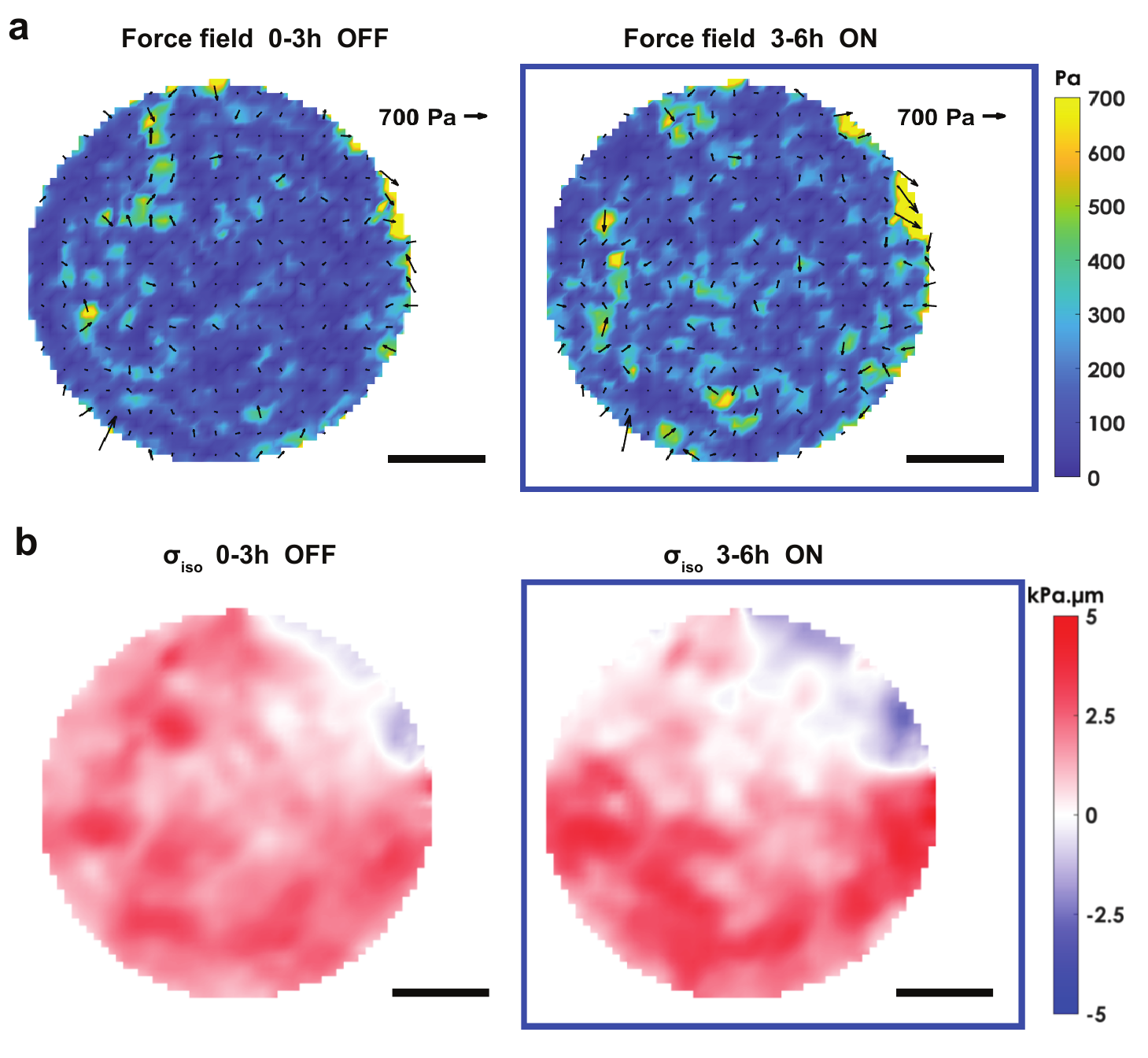}
\caption{\textbf{Traction force field and isotropic stress during OFF/ON experiment.} \textbf{a} Plot of the time-averaged traction forces magnitude 
superimposed with traction forces vector plot obtained by TFM for the first 3 hours (left, OFF) and for the last 3 hours (right, ON, Src is activated). For visibility, only one arrow out of three is drawn. Traction forces are surfacic and thus expressed in~Pa. \textbf{b} Plot of the time-averaged tension $\sigma_{\mathrm{iso}}=(\sigma_{rr} +\sigma_{\theta\theta})/2$ measured by BISM over the first 3 hours (left, OFF) and over the last 3 hours (ON, right, Src is activated). 2d stresses are expressed in kPa$\cdot\upmu$m. Same colony as in Fig.~2, Movie~2. Scale bars: 50~$\upmu$m.}
\label{fig_sup_force_stress}
\end{figure}

\begin{figure}[h!]
\centering
	\includegraphics[width=11cm]{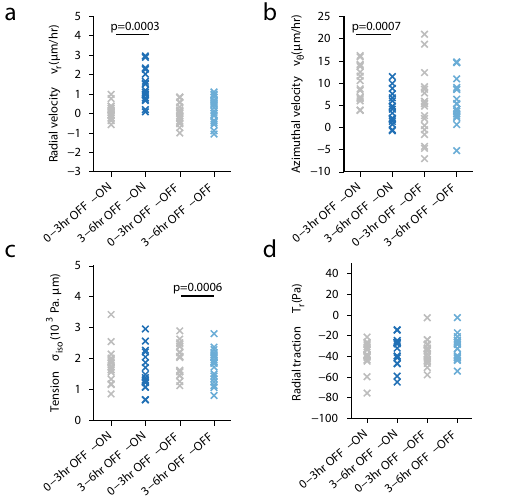}
\caption{\textbf{Spatially-averaged physical fields in the OFF-ON and OFF-OFF experiments}. \textbf{a} Radial velocity component. \textbf{b} Azimuthal velocity component. \textbf{c} Isotropic stress $\sigma_{iso}=(\sigma_{rr}+\sigma_{\theta\theta})/2$. \textbf{d} Radial tractions. All physical variables were averaged on a circle of radius $R$, the confinement radius. Normality was checked  (Shapiro-Wilk test), and T-tests on paired data were performed between the cases 0-3\,hrs OFF-ON and 3-6\,hrs OFF-ON and the cases 0-3\,hrs OFF-OFF and 3-6\,hrs OFF-OFF. Non significant p-values ($p>0.01$) are not shown.} 
\end{figure}

\begin{figure}[h!]
\centering
	\includegraphics[width=14cm]{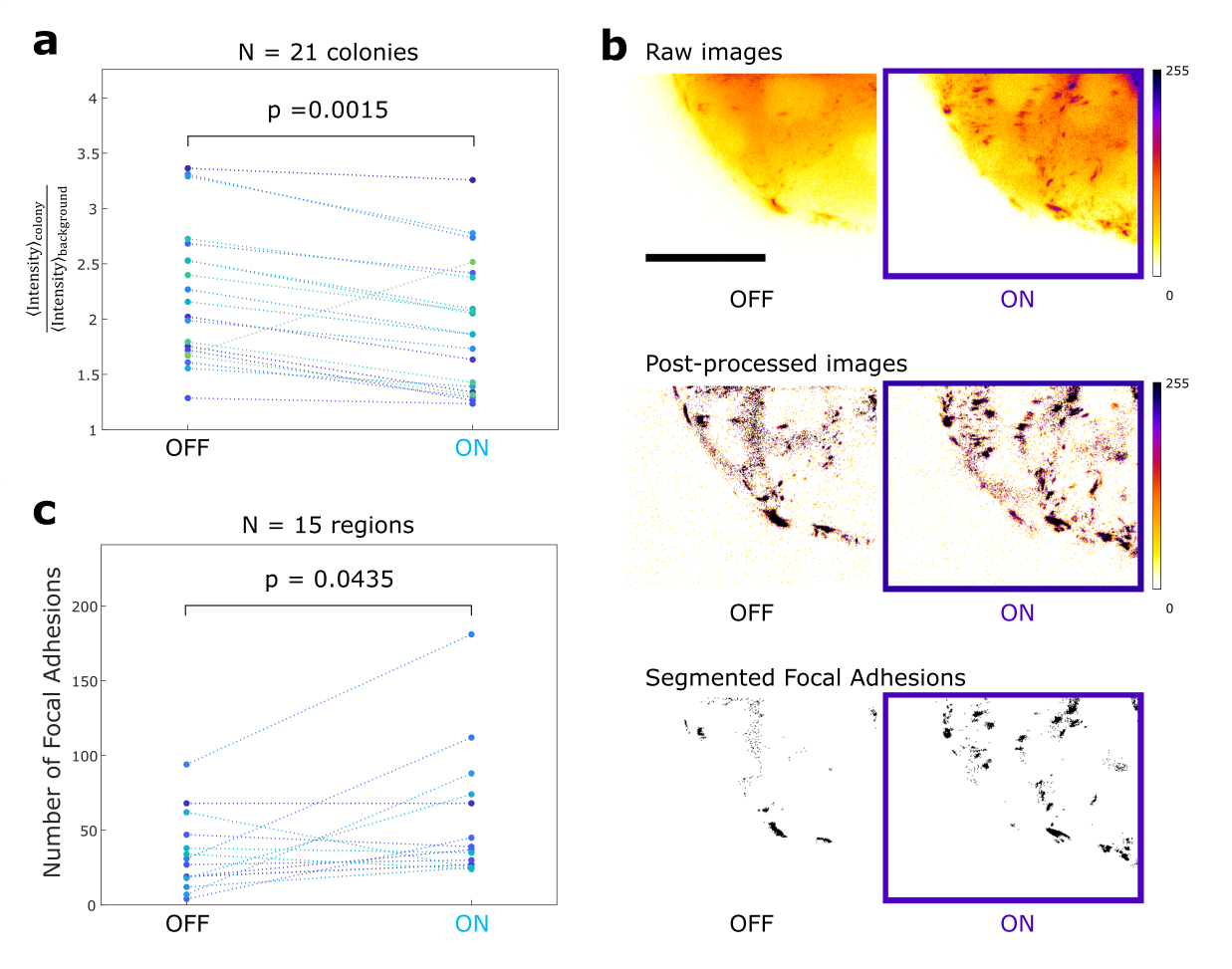}
\caption{\textbf{Quantitative analysis of vinculin distribution.} \textbf{a} Mean intensity of vinculin-iRFP between OFF and ON for all the colonies imaged ($N=21$). The mean intensity inside the circular pattern (colony) was normalized by the mean intensity in a region outside the pattern (background). The tendency to decrease could be associated with the bleaching of the iRFP. \textbf{b} An example of cropped region used in our pipeline to segment focal adhesions (left: cells kept in the dark, right: same region after 1\,hr of Src activation). Top: raw images, middle: images after processing, bottom: segmented focal adhesions. Scale bar: 25\,$\upmu$m. \textbf{c} Comparison of the number of focal adhesions between OFF and ON for $N=15$~analyzed regions. Wilcoxon signed-rank tests were used to check the significance, as the distributions were not normal. See Methods for details about the image analysis steps.}
\end{figure}

\begin{figure}[!h!]
\centering
	\includegraphics[width=9cm]{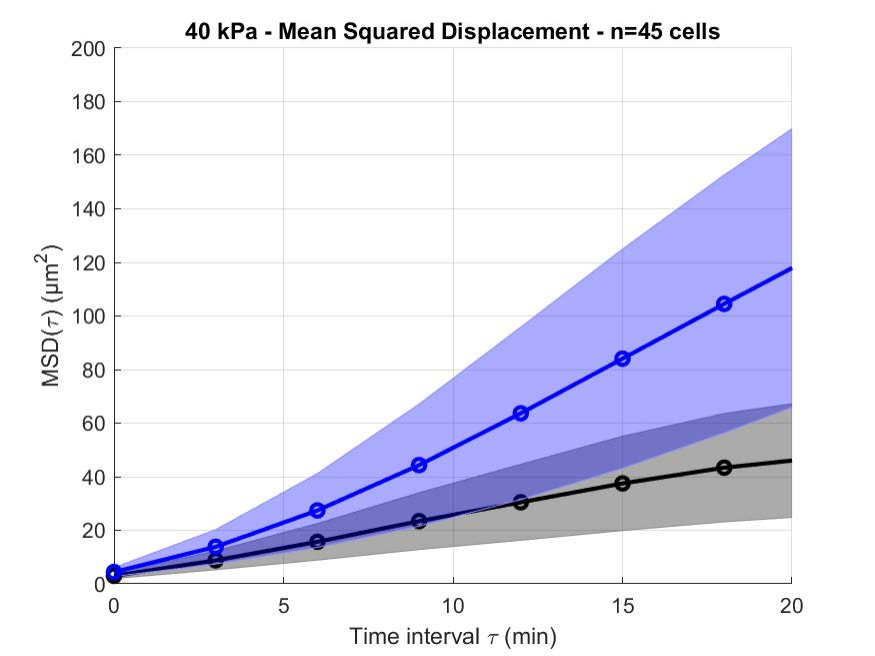}
\caption{\textbf{Mean Square Displacement (MSD) for isolated MDCK optoSrc cells kept in the dark or illuminated with blue light}. $N=45$ isolated cells were subjected to a sequence of 1\,hr in the dark then 1\,hr under blue light, and their center of mass was tracked over time. This gives 45 one-hour tracks for each period. Ensemble-average MSD derived from all tracks OFF (black) and all tracks ON (blue) are shown for the first 20\,min for statistical significance. The error bars represent the 95\% confidence interval for the mean.} 
\label{fig_R2_MSD}
\end{figure}

\begin{figure}[!htb]
\centering
	\includegraphics[width=1.2\linewidth]{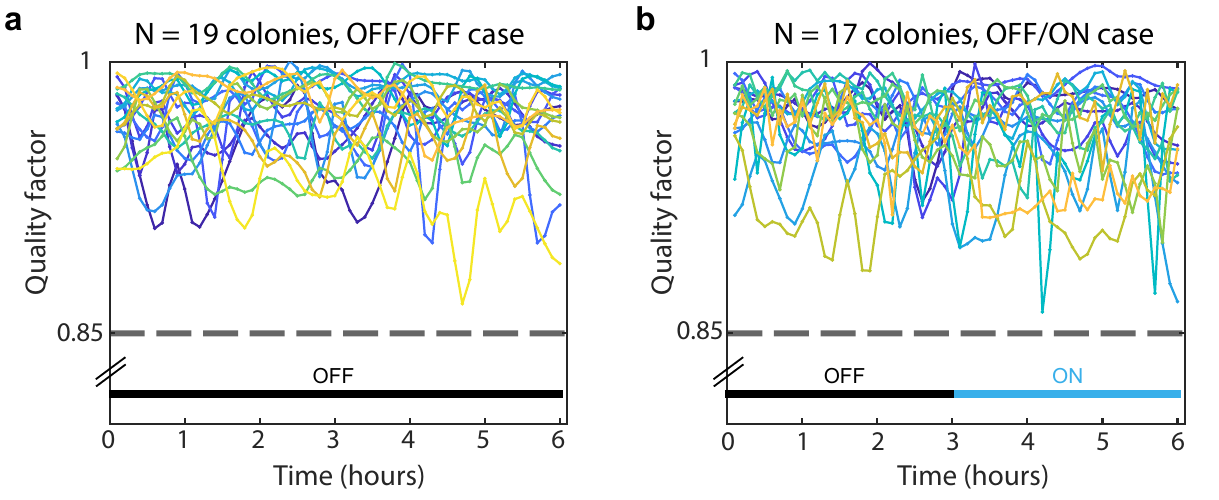}
\caption{\textbf{Quality factor of traction forces reconstruction.} The quality factor is plotted as a function of time for all colonies kept in the dark (left, $N=19$ colonies), or subjected to the transient Src activation (right, $N=17$ colonies). Each colony is represented by a color. The dashed horizontal represents the threshold we have set to estimate that the force reconstruction is acceptable.}
\label{fig_QF}
\end{figure}

\begin{figure}[!htb]
\centering 
\textbf{a.}\includegraphics[scale=0.17]{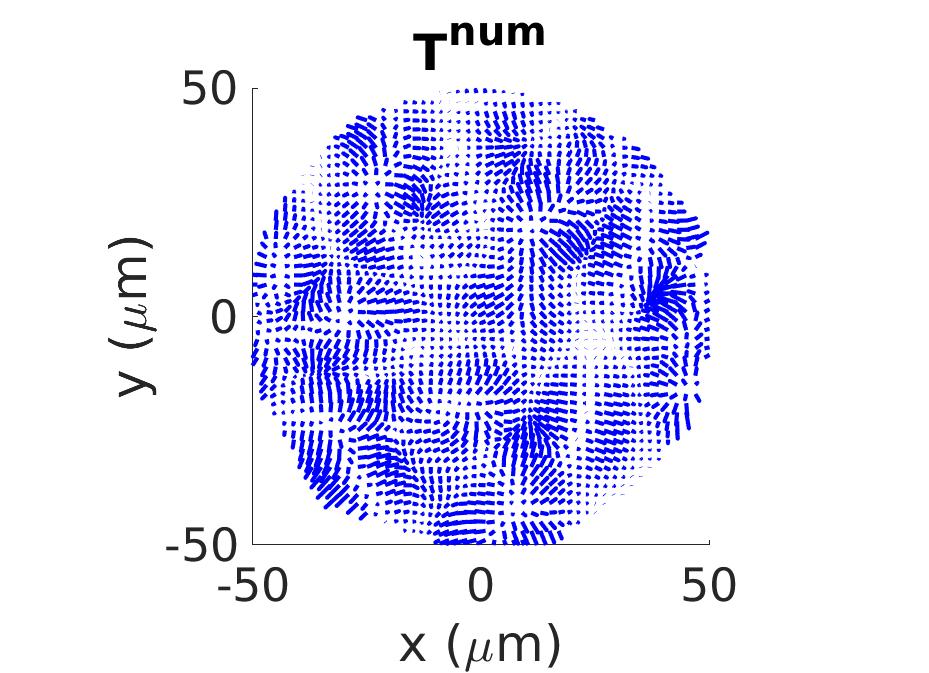}
\hfill
\textbf{d.}\includegraphics[scale=0.17]{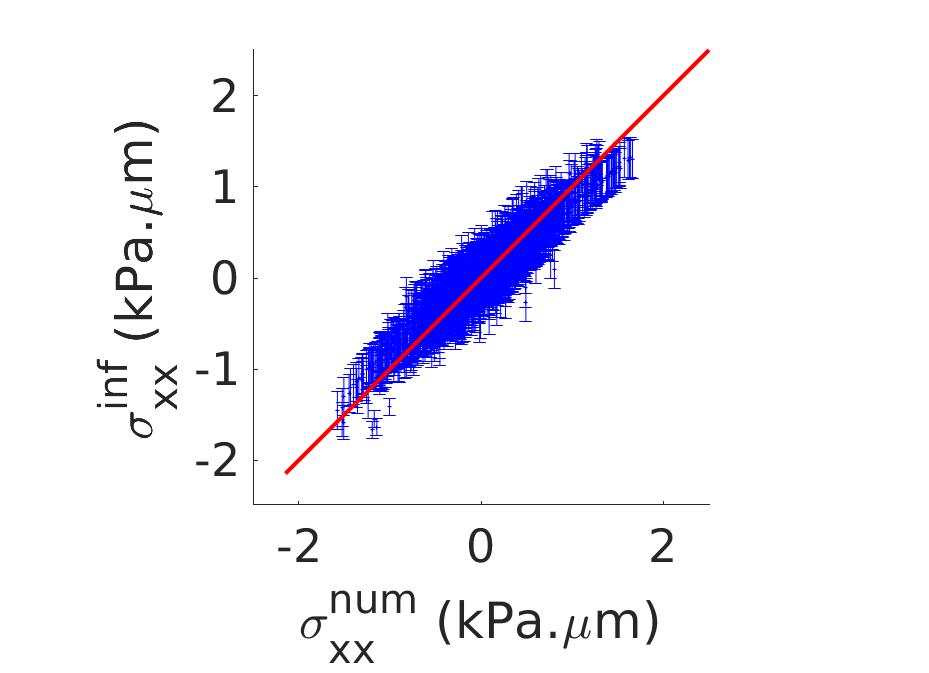}\\
\medskip
\textbf{b.}\includegraphics[scale=0.17]{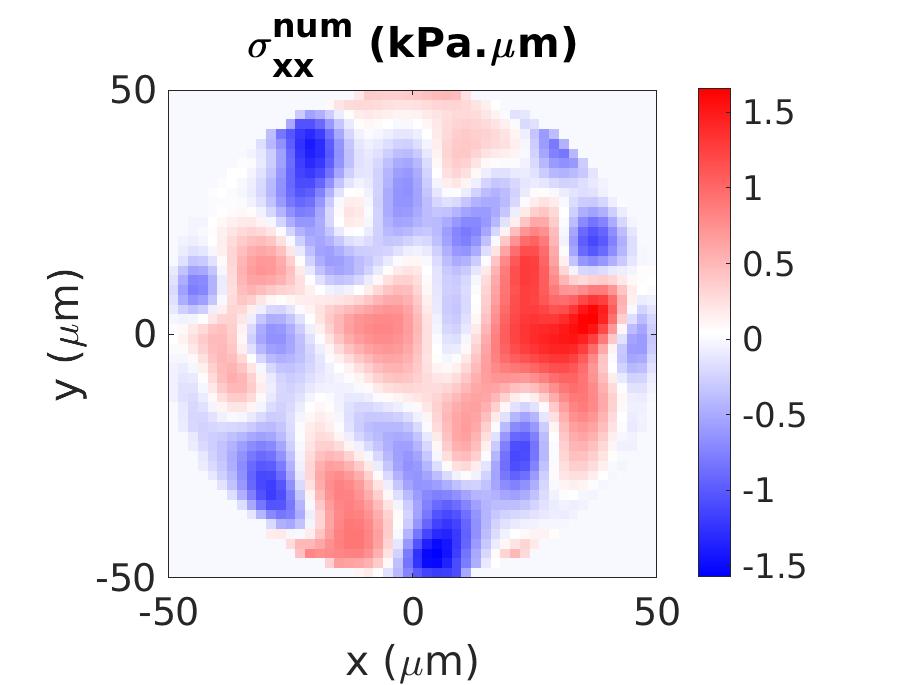}
\hfill
\textbf{e.}\includegraphics[scale=0.17]{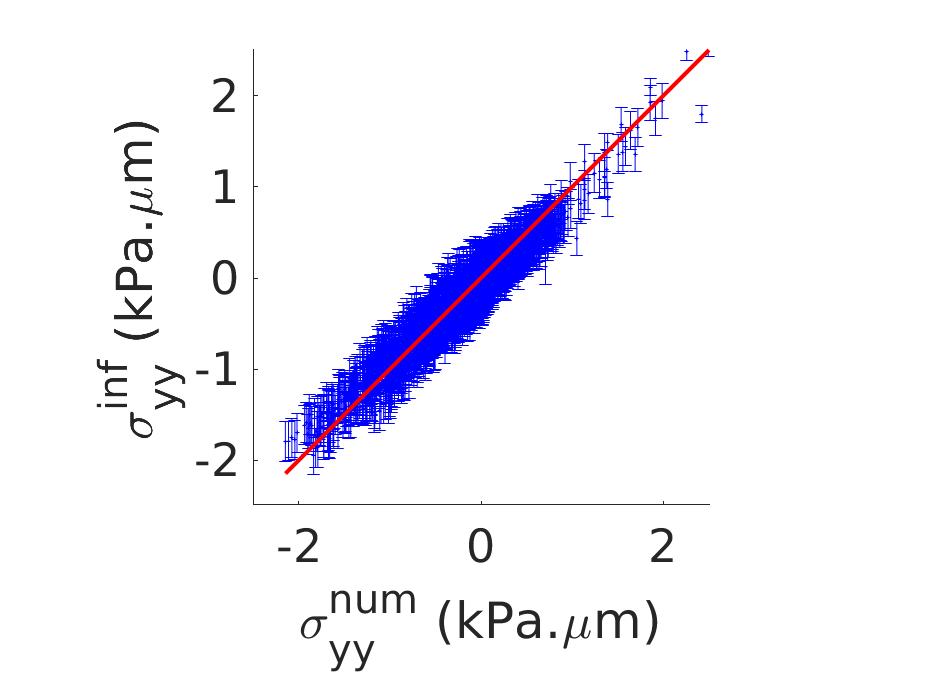}\\
\medskip
\textbf{c.}\includegraphics[scale=0.17]{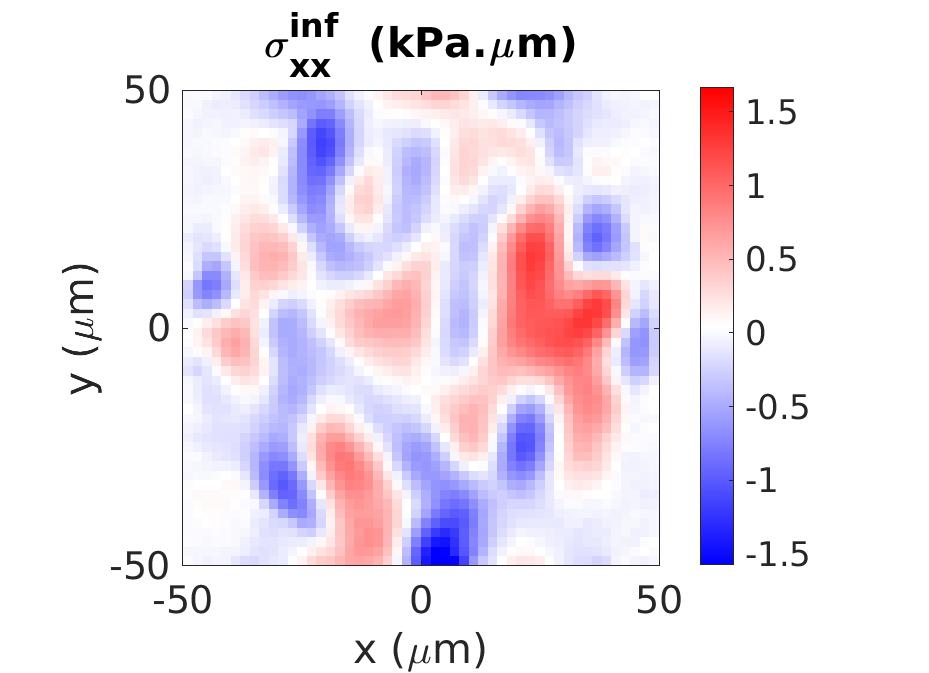}
\hfill
\textbf{f.}\includegraphics[scale=0.17]{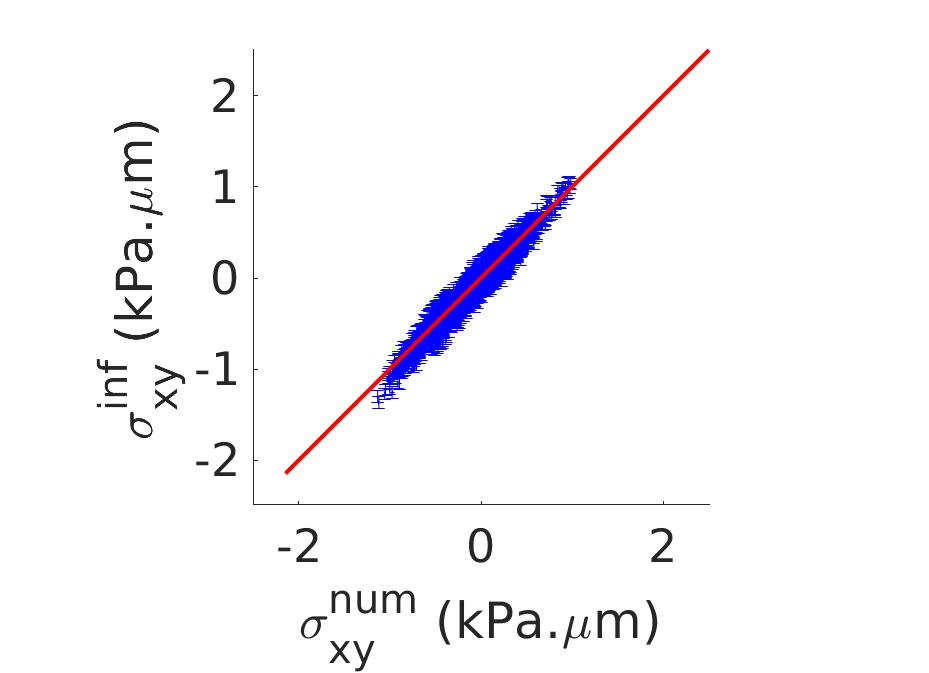}\\
\caption{\label{fig:stress:numerical}
\textbf{Bayesian inversion stress microscopy.} 
\textbf{a.} Simulated traction force field  $\vec{T}^{\mathrm{num}}$.
\textbf{b.} Simulated stress field $x$ component $\sigma_{xx}^{\mathrm{num}}$.
\textbf{c.} Inferred stress field $x$ component $\sigma_{xx}^{\mathrm{inf}}$.
Note the high degree of similarity between images (\textbf{b}) and (\textbf{c}).
\textbf{d-f} Plots of the inferred stress \emph{vs.} the simulated stress 
for each component. Error bars correspond to 
one standard deviation, 
and the red line is the bisector $y = x$. All 2d stresses are given in kPa.$\upmu$m}
\end{figure}

\FloatBarrier
\section{Supporting Movies}
\begin{figure}[!ht]
{\centering
	\includegraphics[width=\linewidth]{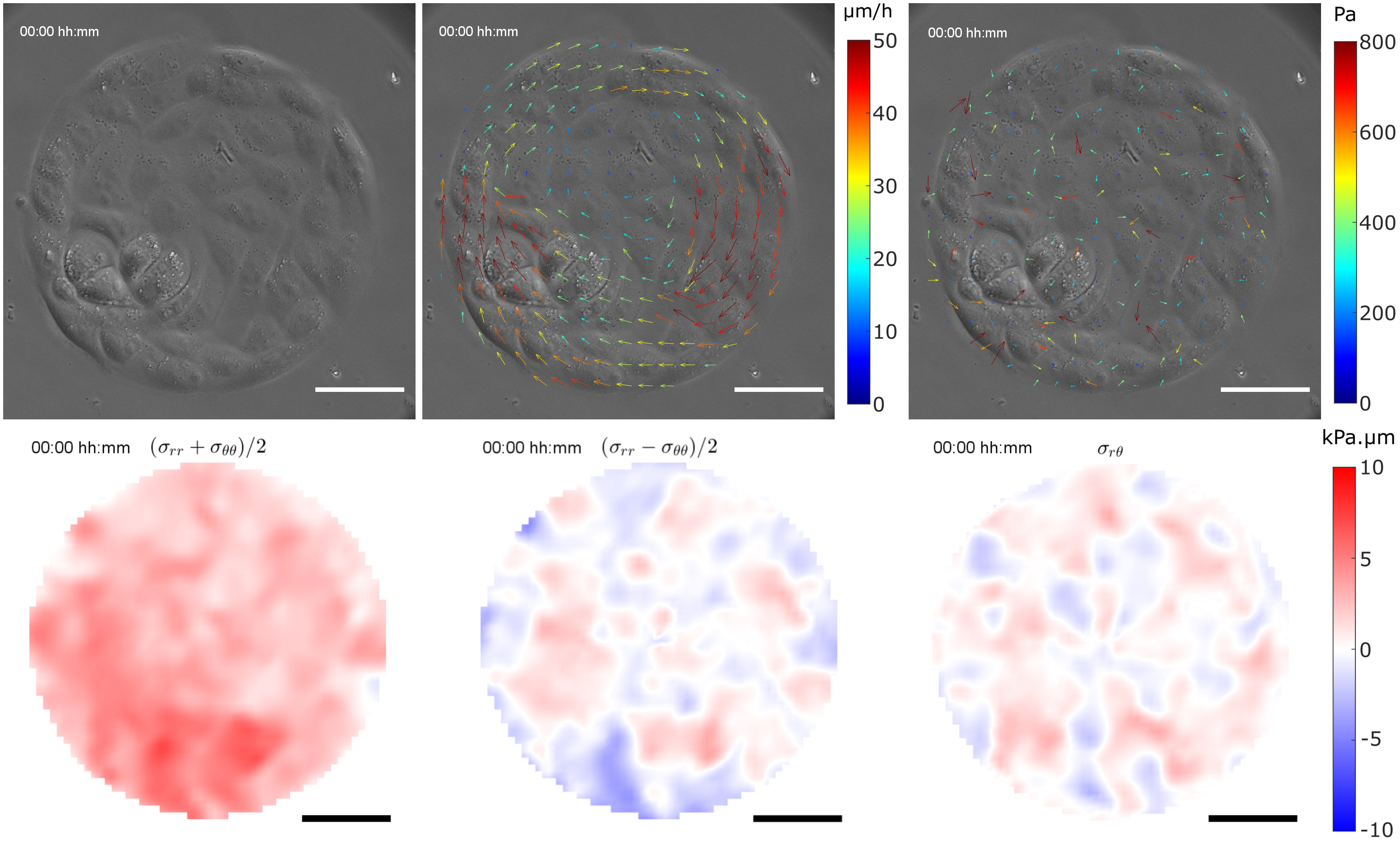}}
	\\[1mm]
\textbf{Mechanical state of confined MDCK optoSrc cells in the dark (first frame of Movie~1).} A typical example of a 6-hour movie of a monolayer of MDCK optoSrc cells, kept in the dark, confined in a 100-µm radius circle, on a polyacrylamide gel of $\sim$40~kPa, displaying collective rotation. In that case, there is no blue stimulation.
1 image every 15~min. Scale bar: 50~$\upmu$m. Same colony as in Fig.~B1.

\textbf{Top-left:} phase-contrast images. \textbf{Top-middle:} velocity field obtained by PIV and superimposed on the phase-contrast images. Color codes for speed, expressed in µm/h. For PIV computation, we used a window size of 128~pixels with 0.5~overlap. Arrows are then interpolated on the TFM grid, and only one of every three is drawn. \textbf{Top-right:} traction forces field obtained by TFM and superimposed on the phase-contrast images. Color codes for traction amplitude, in Pascal. The inter-spacing is 16 pixels, but only one arrow of every two is drawn.

\textbf{Bottom:} Stress components as estimated by BISM. Stresses are given in kPa.$\upmu$m, with the same color code in all panels. \textbf{Left:} tension $(\sigma_{rr}+\sigma_{\theta \theta})/2$. \textbf{Central:} deviatoric component $(\sigma_{rr}-\sigma_{\theta \theta})/2$. \textbf{Right:} shear stress $\sigma_{r \theta}$.
\end{figure}

\begin{figure}[!ht]
{\centering
	\includegraphics[width=\linewidth]{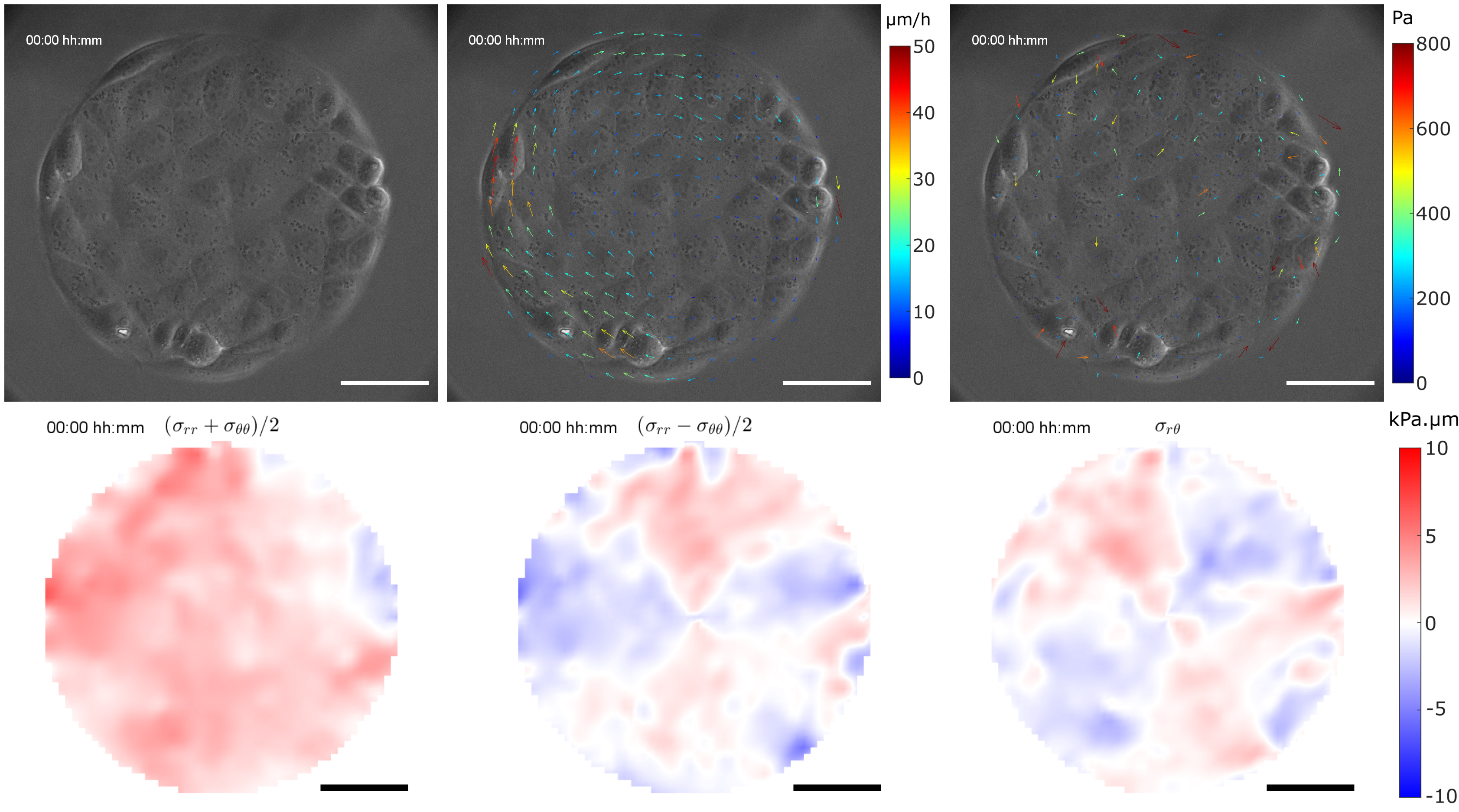}}
		\\[1mm]
\textbf{Mechanical state of confined MDCK optoSrc cells during OFF/ON experiment (first frame of Movie~2).}  A typical example of a 6-hour movie of a monolayer of MDCK optoSrc cells, submitted to the OFF/ON sequence, confined in a 100-µm radius circle, on a polyacrylamide gel of $\sim$40~kPa. During the first 3 hours, the monolayer displays a collective rotation. From $t=3$~h (blue frame), the whole field of view is illuminated with a blue-light pulse every 3~min, the collective rotation is disrupted and the optoSrc cells go outside the pattern. 1 image every 15~min. Scale bar: 50~$\upmu$m. Same colony as in Fig.~2 and Fig.~S5.

\textbf{Top-left:} phase-contrast images. \textbf{Top-middle:} velocity field obtained by PIV and superimposed on the phase-contrast images. Color codes for speed, expressed in µm/h. For PIV computation, we used a window size of 128~pixels with 0.5~overlap. Arrows are then interpolated on the TFM grid, and only one of every three is drawn. \textbf{Top-right:} traction forces field obtained by TFM and superimposed on the phase-contrast images. Color codes for traction amplitude, in Pascal. The inter-spacing is 16 pixels, but only one arrow of every three is drawn.

\textbf{Bottom:} Stress components as estimated by BISM. Stresses are given in kPa.$\mu$m, with the same color code in all panels. \textbf{Left:} tension $(\sigma_{rr}+\sigma_{\theta \theta})/2$. \textbf{Central}: deviatoric component $(\sigma_{rr}-\sigma_{\theta \theta})/2$. \textbf{Right}: shear stress $\sigma_{r \theta}$. 
\end{figure}

\end{document}